\documentclass[12pt]{article}
\usepackage[utf8]{inputenc}
\usepackage{amsmath,amssymb,amsthm,fullpage}
\usepackage[dvipsnames]{xcolor}
\usepackage{graphicx}
\usepackage{subcaption}
\graphicspath{ {./media/} }
\usepackage[normalem]{ulem}
\usepackage{algorithm}
\usepackage{algorithmic}
\usepackage{enumitem}
\usepackage{hyperref}
\usepackage{setspace}
\usepackage{lineno}
\usepackage{authblk}
\usepackage{dsfont}

\title{Bathymetry imposes a global pattern\\of cross-front transport in the Southern Ocean}
\author{Michael C. Denes$^1$}
\author{Shane R. Keating$^{1,2}$}
\author{Gary Froyland$^1$}
\affil{$^1$ School of Mathematics and Statistics,\\ University of New South Wales, Sydney NSW 2052, Australia\\
$^2$ Australian Centre for Excellence in Antarctic Science,\\ University of New South Wales, Sydney, NSW 2052, Australia.\\}
\date{\today}

\begin{document}
\maketitle
\begin{abstract}
The Southern Ocean plays an integral role in the global climate system, exchanging heat, salt, and carbon throughout the major ocean basins via the deep, fast-flowing Antarctic Circumpolar Current.
The Antarctic Circumpolar Current is bounded by spatially and temporally varying fronts that partition distinct water masses.
Locating and quantifying cross-front transport is crucial for understanding global patterns of inter-basin exchange; however, this is challenging because fronts are typically defined by hydrographic properties, such as temperature or salinity, which are subject to external sources and sinks, rather than by properties of the flow itself.
Here we show that, when characterized by material contours that minimize deformation and cross-contour mixing, Southern Ocean fronts exhibit a global pattern of alternating poleward and equatorward transport caused by frontal meandering, which, in turn is influenced by prominent sea-floor obstacles.
These results highlight the importance of bathymetric features in controlling Southern Ocean dynamics and inter-basin exchange.
\end{abstract}

\newpage
\section{Introduction}

The Southern Ocean plays an integral role in the global climate system. It is a key conduit of the global thermohaline circulation \cite{Marshall2012, Morrison2015} and strongly contributes to the exchange of heat, oxygen and carbon between the atmosphere and the deep ocean \cite{Sallee2012}. The Southern Ocean is also crucial to \emph{inter-basin exchange} by connecting the Atlantic, Indian, and Pacific oceans via the Antarctic Circumpolar Current (ACC). The ACC is the world’s strongest current system, deep-reaching and predominantly zonal in nature, driven by a combination of intense westerly winds and steeply tilted density surfaces. The ACC is partitioned by a series of sharp transition zones known as `fronts' that act as barriers to transport between regions of different water mass properties \cite{Chapman2020}. These fronts are the oceanic equivalent of atmospheric fronts, a term coined by Vilhelm Bjerknes to suggest the battlelines drawn between opposing armies.

The location of ACC fronts influence the biological and climate-relevant properties of the Southern Ocean. Frontal regions are known to enhance the upwelling of deep waters and subduction of surface waters \cite{Morrison2015,Stukel2017,Tamsitt2017}, which in turn influence surface chlorophyll concentrations in the Southern Ocean \cite{Sokolov2007}. However, fronts are not impermeable transport barriers. For example, recent genomic analysis and high-resolution ocean modeling has shown that rafting kelps can cross multiple Southern Ocean fronts to beach on the Antarctic coastline \cite{Fraser2018,Fraser2022}. Moreover, observations suggest that the ACC and its fronts are steered and deflected by bathymetry \cite{Moore1999,Sokolov2007}, leading to localized `hotspots' of cross-front transport \cite{Dufour2015,Hughes2010,Patel2019,Rintoul2014,Thompson2012}.

Objective identification and quantification of Southern Ocean cross-front transport is critical for understanding global patterns of inter-basin exchange \cite{Chapman2020}. This is a challenging task, however, as most definitions of oceanic fronts are traditionally based upon water-mass criteria --- for example, a particular value of temperature at depth \cite{Belkin1996,Orsi1995} --- that are subject to external sources and sinks. Such definitions were a practical necessity due to the paucity of early Southern Ocean observations and datasets. However, in recent decades, observations from satellites, ARGO floats, and ocean drifters have greatly added to our understanding of the Southern Ocean, the ACC, and its fronts. Satellite imagery in particular has revealed the dynamic and filamentary nature of Southern Ocean fronts \cite{Chapman2017}. This has led to a range of new front definitions, sometimes leading to conflicting conclusions \cite{Chapman2020}.

\begin{figure}
\centering \includegraphics[width=0.8\linewidth]{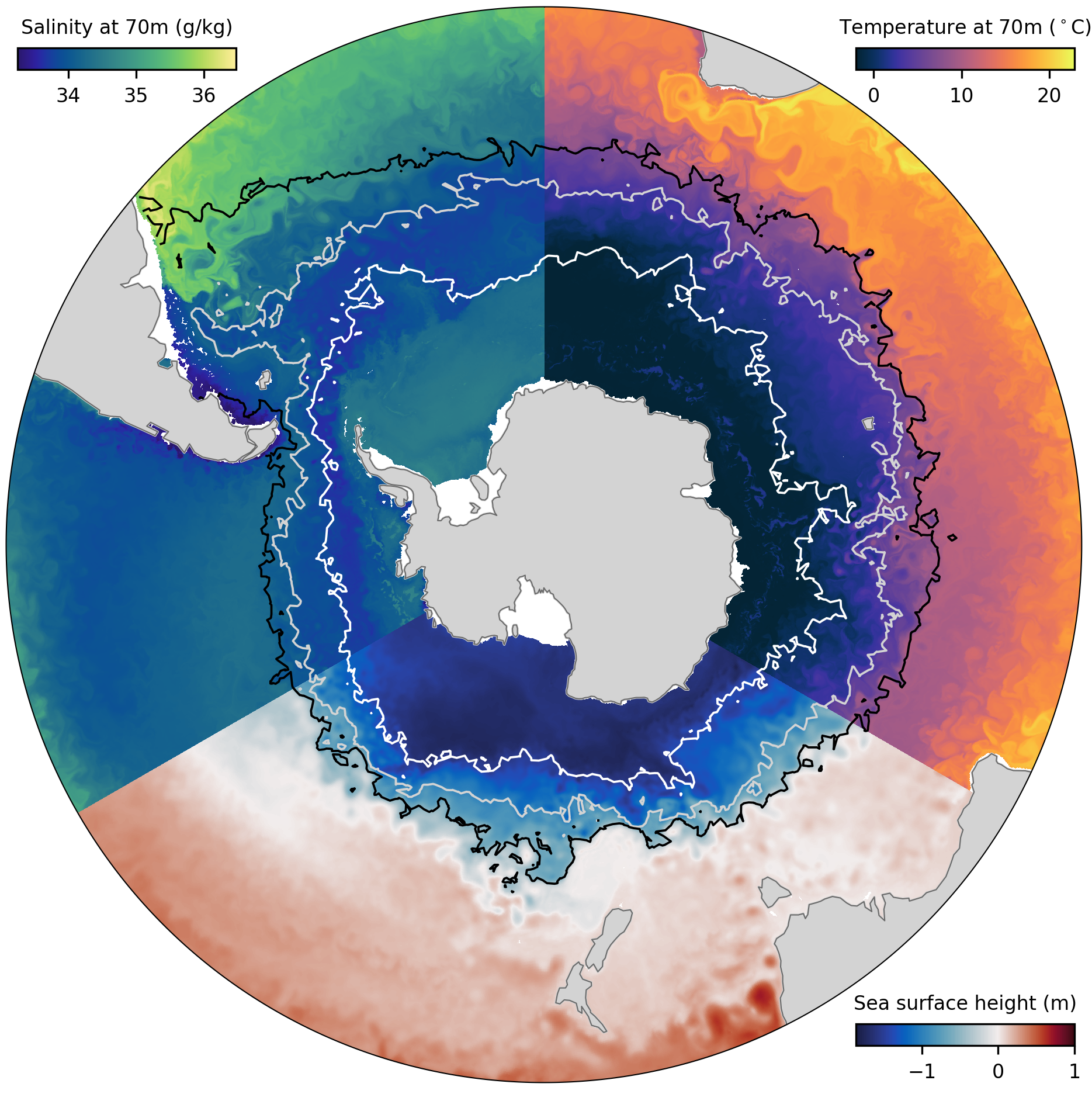}
\caption{\textbf{Southern Ocean fronts identified as flow-following material contours that minimize cross-front transport.} Three concentric circumpolar fronts are identified as material contours that move with the flow while minimizing deformation and mixing over a 540-day time window (Methods). Each front is plotted at the center of the time window (day 270).
The fronts are computed using simulated particle trajectories at fixed depth (70m)  calculated from a 0.1$^\circ$ global eddy-resolving ocean model. Background colors in each sector show a snapshot (on day 270) of daily averaged practical salinity at 70m (top left), conservative temperature at 70m (top right), and effective sea surface height (bottom). The identified fronts partition the Southern Ocean into sub-domains of different water mass properties, often aligning with steep gradients in the underlying physical fields.}
\label{fig:fronts_fig1}
\end{figure}

Here, we take an alternative and as-yet unexplored approach to investigate cross-front transport that is based on the properties of the flow alone. We identify Southern Ocean fronts as circumpolar material contours that are minimally stretched and distorted by the flow. Because diffusive mixing across the front is proportional to the evolved contour length \cite{ShuckburghHaynes2003}, these contours will act as transport barriers that minimize water mass transport across the front. In Figure \ref{fig:fronts_fig1} we plot three such contours, overlaid on a snapshot of the daily averaged conservative temperature, practical salinity, and sea surface height (a proxy for pressure) computed from a high resolution ocean model. Although the material contours are defined by the flow alone, it can be seen that they nonetheless trace fronts of these hydrographic properties.

The fronts shown in Figure \ref{fig:fronts_fig1} are identified as minimally mixing contours over a 540-day window. The fronts partition the Southern Ocean into two materially coherent sub-domains over this time period. By considering a sliding-in-time sequence of windows, we can locate and quantify cross-front transport by identifying regions of fluid that lie on one side of the front in one window and on the other side of the front in the next. In this work, we show a clear, global pattern of alternating poleward and equatorward cross-front transport in the Southern ocean, distinct from Ekman and wind-driven transport. We demonstrate that the global pattern of cross-front transport is caused by frontal meandering in regions downstream of prominent bathymetric obstacles. These results highlight the importance of the Southern Ocean bathymetry in controlling frontal dynamics and provide a new framework for understanding inter-basin exchange of climate-relevant water properties.

\newpage
\section{Results}

\subsection*{Southern Ocean fronts as boundaries of finite-time coherent sets}\label{sec:FTCS}

Lagrangian coherent structures are fluid structures that move with the flow and remain materially coherent for a given window of time \cite{Froyland2010,Haller2012,Allshouse2015,Hadjighasem2017}. They play an important role in transport and mixing and have seen wide use in oceanographic applications, in particular mesoscale ocean eddy identification \cite{Beron-Vera2008,Froyland2012,Froyland2015c,Haller2016,Abernathey2018,Denes2022}. In the dynamical systems literature, a range of techniques has been developed to detect Lagrangian coherent structures (see \cite{Hadjighasem2017} for a recent critical review). In the context of oceanic fronts, several studies identify fronts as ridges in the finite-time Lyapunov exponent \cite{Abraham2002} or finite-size Lyapunov exponent field \cite{Bettencourt2015, dOvidio2015, Lehahn2007}. However, in general, it is difficult to relate the strength of finite-time/size Lyapunov exponent field to cross-front transport as these fields describe regions that experience large trajectory separation and provide no information about material transport. Applications of other Lagrangian methods to Southern Ocean fronts have thus far been unexplored.

Here, we identify Southern Ocean fronts as the boundaries of \emph{finite-time coherent sets}, regions of phase space that minimally mix with the surrounding phase space over a finite window of time \cite{Froyland2010,Froyland2013}. The boundaries of these regions are Lagrangian coherent structures that best resist distortion and filamentation and thereby minimize mixing by small-scale diffusion. As such, finite-time coherent sets provide an objective mathematical framework for identifying transport barriers in phase space. In the ocean context, a finite-time coherent set describes a region of fluid bounded by a material curve or surface that moves with the flow. Here, we will take a geometric approach to identify finite-time coherent sets by considering the dynamic isoperimetric problem \cite{Froyland2015a,Froyland2018} which seeks sets with minimal evolving boundary size relative to enclosed volume. The numerical approach described in (Methods) computes finite-time coherent sets as sub-level sets of the leading eigenfunction of the dynamic Laplace operator \cite{Froyland2015a, Froyland2018}. Our approach extends the method developed in \cite{Froyland2018,Abernathey2022,Denes2022} for use with Lagrangian oceanographic datasets in polar regions.

In this study, we are interested in lateral transport across a front, so our focus will be on a two-dimensional domain representing a specific depth level of the global ocean south of $30^\circ$S. We denote by $\mathcal{M}$ this region of ocean at 70m depth, with all land removed. The domain is restricted to the 70m depth level in order to remove the wind-driven transport effects within the Ekman layer. While we have limited our study to lateral transport at a fixed depth, extensions of the method to higher dimensions with appropriate modifications can be used to investigate the three-dimensional structure of Southern Ocean fronts and both lateral and vertical transport across fronts.

We are interested in circumpolar fronts, and therefore seek sets whose boundaries do not intersect the coastline of Antarctica. On the other hand, we do not wish to impose any restrictions regarding intersection with other continents.
To achieve this, we impose Dirichlet boundary conditions \cite{Froyland2018,Abernathey2022,Denes2022} on the Antarctic coastline, while leaving natural (Neumann) boundary conditions on the remainder of the domain boundary.
This work represents the first mixed-boundary computation of finite-time coherent sets, motivated by the annular topology of the Southern Ocean around Antarctica.

Denote by $\text{int}(\mathcal{M})$ the interior of $\mathcal{M}$, i.e., the ocean south of $30^\circ$S not including coastlines and the northern open ocean boundary at $30^\circ$S.
Let the set $A\subset \text{int}(\mathcal{M})$ be a region of ocean bounded by a material contour and/or continent that completely surrounds Antarctica. Define a window of time $W(0,T) = [-T/2, T/2]$ of length $T>0$ centered at time $t=0$. The set $A$ that minimizes the ratio of its average evolving boundary length to enclosed area over $W(0,T)$ defines a finite-time coherent set on this window. Such a set $A$ represents a region of ocean surrounding Antarctica that minimally mixes with more northern waters. The northern boundary of $A$, which we denote by $\partial A$, identifies a material contour that best resists distortion and filamentation over $W(0,T)$. Thus, $\partial A$ defines a circumpolar transport barrier --- i.e., a Southern Ocean front --- over the specified time window.

We plot three concentric Southern Ocean fronts in Figure \ref{fig:fronts_fig1}, labelled (from south to north) $\partial A_{1}$ (southern), $\partial A_{2}$ (middle), and $\partial A_{3}$ (northern), that represent the boundaries of nested coherent sets $A_1$, $A_2$, $A_3$ calculated over a window of length $T = 540$ days. The fronts identified using this method are determined solely by the flow itself: they contain no information about hydrographic properties such as temperature, salinity, or sea-surface height, except insofar as these quantities influence the velocity field. Nonetheless, as shown in Figure \ref{fig:fronts_fig1}, the minimally distorted curves closely reproduce gradients or contours of water-mass properties that are traditionally used to identify fronts. Indeed, the boundaries  $\partial A_{1}, \partial A_{2},$ and $\partial A_{3}$ were chosen in part because they resemble three widely used circumpolar fronts identified from hydrographic data and water-mass criteria, namely, the Polar Front, the Sub-Antarctic Front, and the Subtropical Front, respectively \cite{Orsi1995}; however, we make no claim that the boundaries define these hydrographic fronts. In Figure \ref{fig:fronts_ftcs_diagram} in the Supplementary Material we illustrate how one of the material contours, $\partial A_2$, resists distortion and filamentation, compared to three latitudinal rings, as they are advected by the flow. In subsequent tests, we found that the method is insensitive to the choice of window length $T$, provided the window is sufficiently long to adequately capture the dynamics of the flow (see Figure \ref{fig:telescoping_eigenfunctions} in the Supplementary Material and \cite{Denes2022} for a detailed discussion on the choice of window length). The method was also able to identify the fronts in both global and regional domains, with appropriate modifications to the boundary conditions (see Supplementary Material).

The method described above identifies fronts as minimally distorted material contours over a prescribed time window $W(0,T)$. By incrementally changing the window center time, we generate a \emph{sliding sequence} of windows $W(t_c,T) = [t_c-T/2, t_c+T/2]$ parameterized by the center times $t_c \in [t_{\min}, t_{\max}]$ for selected $t_{\min} < t_{\max}$. From these windows, we can generate a time series of front locations, anchored to the center time $t_c$, composed of a sequence of material contours from the center of each window. To `match' fronts from one window to the next we use an equal-area constraint, that is, the area enclosed by the front in one window must equal that enclosed by the front in the following window. An animation of the time series of fronts identified from a sequence of sliding windows can be found in the Supplementary Material.

Although each front represents a minimally distorted material contour over the window on which it is defined, the time series of front locations does \emph{not} track the evolution of a material contour. That is, a front anchored on day $t$ and evolved forwards by $s$ days will not correspond to a front anchored on day $t+s$. In fact, the difference between these contours, and the coherent sets they enclose, represents the transport across the circumpolar front. Specifically, a region that is on the poleward side of the front anchored at time $t$ but on the equatorward side of the front anchored at time $t+s$ pulled back by $s$ days is considered to have `exited' the coherent set. Likewise, a region that is on the poleward side of the front anchored at time $t+s$ but is on the equatorward side of the front anchored at time $t$ pushed forward by $s$ days is considered to have `entered' the coherent set. Because the fronts we identify are predominantly zonal we will speak without ambiguity of `entering' material as southward transport and `exiting' material as northward transport.

\begin{figure} 
\centering \includegraphics[width=0.95\linewidth]{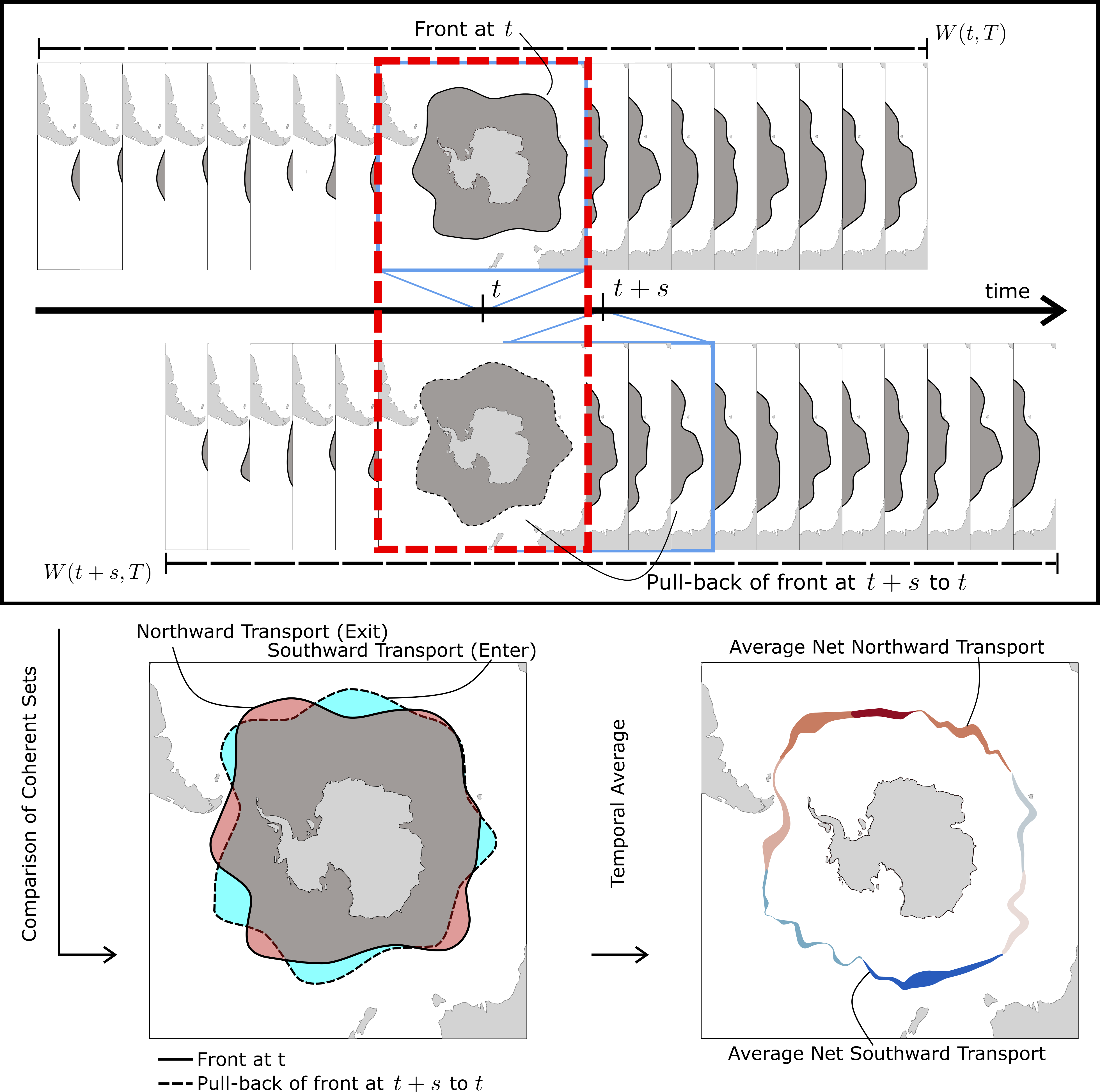}
\caption{\textbf{Locating and quantifying cross-front transport.} A schematic in three parts, which shows the process of locating and quantifying the cross-front transport along a front. The top panel shows two fronts identified from windows $W(t,T)$ and $W(t+s,T)$, anchored at times $t$ and $t+s$ respectively, highlighted by blue boxes. To identify cross-front transport, the front at time $t+s$ is pulled back by $s$ days, so that both material curves are anchored at time $t$, shown in the red dashed box. The regions contained by these two fronts are compared in the bottom left panel, which identifies regions of northward (exit) cross-front transport (red regions) and southward (entry) cross-front transport (blue regions). A time series of northward and southward transport regions is constructed by repeating this process over a range of center times. A temporal average of these regions is taken. The average northward, southward, net, and total cross-front transport spatial maps can then be analyzed in order to determine if any pattern of cross-front transport exists, as illustrated in the bottom right panel.}
\label{fig:fronts_fig2}
\end{figure}

We sketch this process in Figure \ref{fig:fronts_fig2}. We start by identifying the front anchored at time $t$ for the window $W(t,T)$, and the $s$ day pull-back (or backward evolution) of the front anchored at time $t+s$ for the window $W(t+s,T)$. These fronts are identified over two different (but overlapping) windows, but the fronts are both instantiated at the same time $t$. We then compare the corresponding regions of ocean bounded by these fronts, and the Antarctic coastline, in order to identify exit (northward transport) regions and entry (southward transport) regions. By repeating this process for all center times and taking a temporal average over each region of ocean, we can quantify the average amount of northward, southward, and net cross-front transport for each location and identify spatial patterns of cross-front transport.

\subsection*{A global pattern of cross-front transport}

The process described above (and sketched in Figure \ref{fig:fronts_fig2}) was used to calculate transport across a circumpolar front in a global, high-resolution numerical ocean model. Here, we focus on the middle front identified in Figure \ref{fig:fronts_fig1}, $\partial A_{2}$, as it resembles and shows characteristics of the Sub-Antarctic Front (SAF) and passes through several regions of major bathymetric features known to be important for cross-front transport \cite{Dufour2015, Thompson2012}. As an initial estimate for the window length, we use $T=540$ days, which roughly equates to the time it takes a particle to circumnavigate Antarctica along a line of latitude at $55^\circ$S assuming an average speed of $0.5$ m/s. For comparison, we computed fronts for several window lengths, $T= \left\{180,360,540,720\right\}$ for center time $t_c=365$, and saw no appreciable difference in the structure of the $T=540$ and $T=720$ eigenfunctions (see Figure \ref{fig:telescoping_eigenfunctions} in the supplementary material). This gives us confidence that $T=540$ is an appropriate choice of window length.

We use six years of daily particle trajectory positions, which are computed using daily averaged velocities at 70 m depth from the ACCESS-OM2 ocean-sea ice model \cite{Kiss2020} ($0.1^\circ$ horizontal spacing). We identify daily front positions for windows of length $T=540$ days and center time $t_c=270$ (model date 27/09/2008) through $t_c=1920$ (model date 04/04/2013). We compare front positions from windows shifted by $s = 30$ days in order to construct a time series of entry and exit regions. We then calculate the area of the entry and exit regions and divide by $30$ days to obtain a southward and northward area flow rate in units of km$^2$ per day for the period $t=270$ through $t=1890$. Next, we bin the area flow rate on a uniform $0.1^\circ \times 0.1^\circ$ longitude-latitude grid for each center time and time average over the window center times (nearly 4.5 years from the center of the first window to the center of the last window). To calculate the average northward and southward area fluxes (in km/day) we normalize the area flow rate by the bin width (in km) along the bottom edge of the box. This flux can be interpreted as the average distance a parcel of water travels per day once it crosses the front.

\begin{figure}[H]
\centering \includegraphics[width=\linewidth]{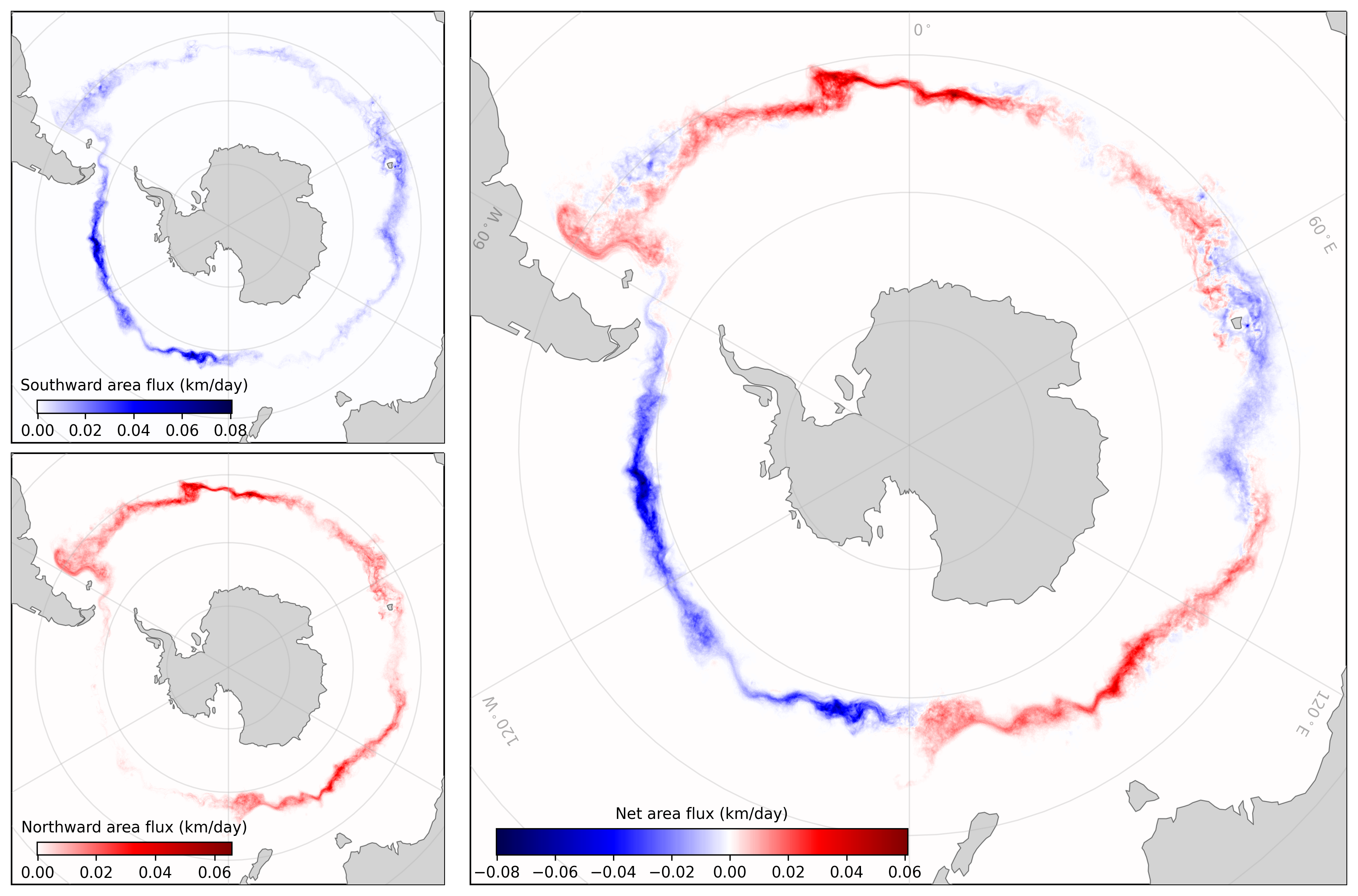}
\caption{\textbf{A global pattern of cross-front transport.} Southern Ocean fronts exhibit a strong geographic pattern of cross-front transport. A 4.5-year sequence of daily front locations is constructed by tracking the middle front identified from Figure \ref{fig:fronts_fig1}. For each center time, the northward (red) and southward (blue) cross-front transport regions are identified. A temporal average of these regions normalized by the contour length is computed on a $0.1^\circ \times 0.1^\circ$ uniform grid in order to compute an area flux. The average southward cross-front area flux is plotted in the top left panel, and the average northward cross-front area flux in the bottom left panel. The average net cross-front transport is computed as the difference of these two fields, show in the right panel. A distinctive global pattern of net cross-front transport emerges, where the net cross-front transport transitions from northward to southward, or southward to northward, downstream of prominent bathymetric obstacles. After each transition we see particularly large amounts of total cross-front transport.}
\label{fig:fronts_fig3}
\end{figure}

We plot the average daily northward, southward, and net cross-front flux in Figure \ref{fig:fronts_fig3}. A distinct global pattern of cross-front transport emerges: northward cross-front transport is confined to the South Atlantic, West Indian, and East Indian/Tasman regions, while southward cross-front transport is confined to the South Pacific and Central Indian/Kerguelen regions, with some southward cross-front transport emerging in the South Atlantic. The locations of particularly large total cross-front transport are consistent with \cite{Dufour2015, Thompson2012}. However, a surprising feature of Figure \ref{fig:fronts_fig3} is that most frontal regions are dominated by cross-front transport in one direction, i.e. northward or southward. Indeed, only two regions across the domain showing mixed northward and southward cross-front transport. These occur downstream of the Drake passage and downstream of the Southwest Indian ridge, before the front reaches the Kerguelen plateau.

Across most of the Southern Ocean, we observe a clear alternating pattern of net cross-front transport that switches from northward to southward and back at distinct transition zones separated by basin scales (thousands of kilometers). Specifically, in Figure \ref{fig:fronts_fig3} we identify four transition zones between net southward (blue) or northward (red) cross-front transport regimes. Moving eastward (clockwise) from $0^\circ$ degrees (``noon'') in Figure \ref{fig:fronts_fig3}, right panel, net cross-front transport in the South Atlantic is predominantly northward until the first transition region when the front reaches the Kerguelen plateau ($\sim 65^\circ$E). Downstream of this transition zone, net cross-front transport is southward until the second transition zone occurs on the Southeast Indian ridge ($\sim 100^\circ$E) where there is a transition to net northward cross-front transport. The third transition zone occurs downstream of the Campbell plateau ($\sim 180^\circ$E), where a transition from net northward cross-front transport to net southward cross-front transport occurs over the width of the South Pacific. Finally, a fourth transition occurs downstream of the Drake passage ($\sim 60^\circ$W), resulting in a transition from net southward cross-front transport in the South Pacific to net northward cross-front transport in the South Atlantic.

We note that, while our computations have been performed on an almost 4.5-year-long time series of exit and entry regions, a qualitatively similar global pattern is apparent for shorter observation periods of one to three years (all starting from $t=270$). Moreover, while the net area flux along the entire front is zero (by definition, see Methods), this does not explain the clear global pattern of alternating northward and southward cross-front transport punctuated by transition zones that occur downstream of bathymetry.

\subsection*{Bathymetry leads to cross-front transport controlled by meandering}

We now turn our attention to the drivers of the observed global pattern of cross-front transport. The ACC is a deep-reaching current, extending from the surface to depths down to 4000m, and is strongly influenced by bathymetric features like plateaus, fracture zones, ridges, and sea mounts \cite{Moore1999,Sokolov2007,Thompson2010,Graham2012}. These bathymetric obstacles are important to the mean circulation of the Southern Ocean, deflecting or diverting the path of the ACC away from its typically zonal path \cite{Marshall1995,Moore1999,Sokolov2007,Thompson2010} in order to conserve potential vorticity \cite{VallisMaltrud1993,Gille1997a,Gille1997b}. A number of investigations of the dynamics of Southern Ocean jets and fronts near bathymetric features found an increased variability in their location downstream of prominent bathymetry \cite{Sokolov2009b,Graham2012,Chapman2014, Shao2015}. Bathymetry induces standing meanders in the ACC jets \cite{Thompson2014}, which reduce the transport barrier effect and enhance eddy-induced cross-jet transport \cite{Abernathey2014, Barthel2022}.

\begin{figure}
\centering
\includegraphics[width=\linewidth]{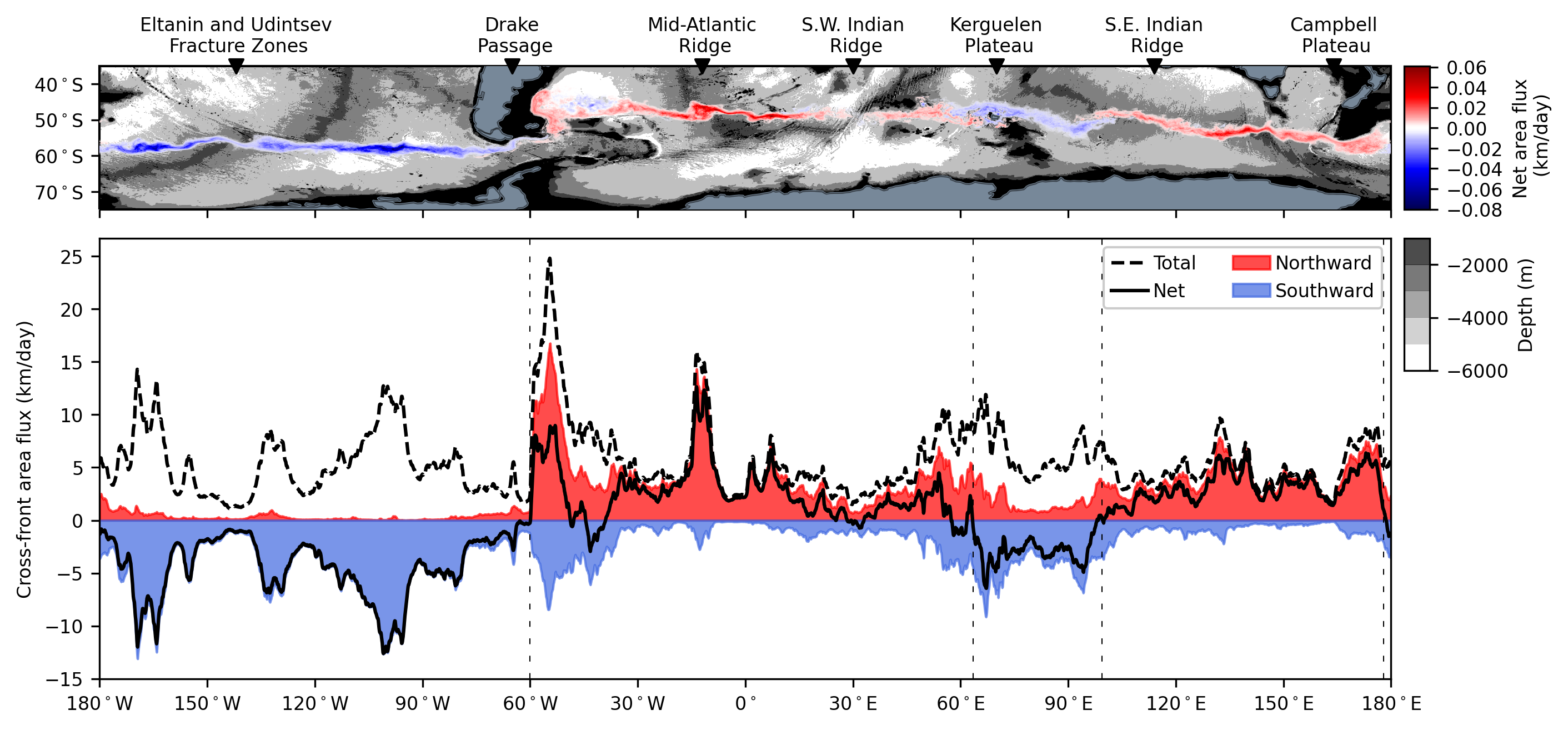}
\caption{\textbf{Bathymetry leads to cross-front transport.} Regions of increased cross-front transport occur downstream of prominent bathymetric obstacles. The time-averaged net cross-front transport pattern shown in Figure \ref{fig:fronts_fig3} is shown over the model bathymetry (in grayscale) in the top panel. Small values have been made transparent. The time-averaged southward, northward, net, and total cross-front transport amounts are plotted in the bottom panel as a function of longitude. Vertical black dashed lines indicate the locations of the four transition zones. The peaks in cross-front transport typically occur downstream of prominent bathymetric obstacles. We have labeled a few of the most prominent bathymetric features found in the Southern Ocean.}
\label{fig:fronts_fig4}
\end{figure}

To investigate the influence of bathymetry on cross-front transport, we plot in Figure \ref{fig:fronts_fig4} the average net cross-front transport (as was shown in Figure \ref{fig:fronts_fig3}) in an equidistant cylindrical projection over the model bathymetry. To complement this, we plot the average northward, southward, net (northward minus southward), and total (northward plus southward) cross-front transport per $0.1^\circ$ of longitude, that is, the sum in the meridional axis. We observe that strong cross-front transport is confined to key regions downstream of prominent bathymetric obstacles. Southward transport (shown in blue in Figure \ref{fig:fronts_fig4}) dominates in two regions: (i) the South Pacific, where peak southward transport occurs downstream of the Campbell plateau, ($\sim 165^\circ$W),
and downstream of the Eltanin and Udintsev fracture zones; ($\sim 130^\circ$W),
and (ii) the central Indian Ocean, where peak southward transport occurs on and downstream of the Kerguelen plateau. ($\sim 65^\circ$E and $\sim 90^\circ$E, respectively).
The majority of northward cross-front transport (shown in red in Figure \ref{fig:fronts_fig4}) is found in three regions: (i) the South Atlantic region, where maxima in northward cross-front transport occur downstream of the Drake passage and Nova Scotia ridge, and as the front passes over the Mid-Atlantic ridge.
(ii) the West Indian region, where the peak in northward transport occurs downstream of the Southwest Indian ridge near the Crozet Plateau ($\sim 110^\circ$E);
(iii) the East Indian/Tasman region, where the peaks in northward transport occur downstream of the Southeast Indian ridge once the front has passed the Australian-Antarctic discordance ($\sim 135^\circ$E),
and immediately downstream of the Macquarie ridge and Campbell plateau ($\sim 175^\circ$E).

\begin{figure}
\centering
\includegraphics[width=\linewidth]{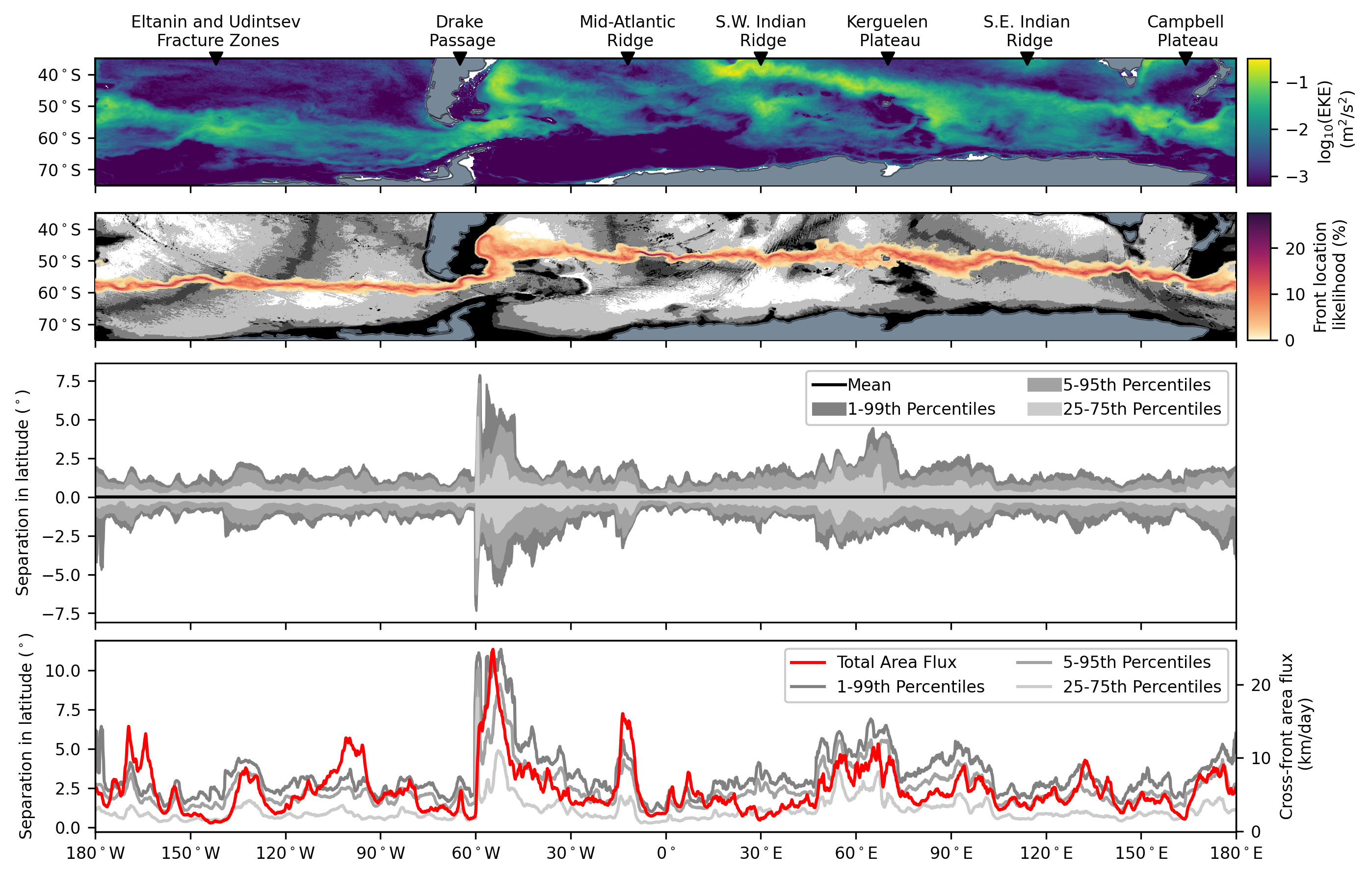}
\caption{\textbf{Locations of meandering correspond to strong cross-front transport.} Increased cross-front transport is directly linked to meandering of the fronts, which occurs downstream of prominent bathymetric obstacles. This meandering occurs both in and downstream of regions of enhanced EKE. The EKE field is computed at 70m depth from 6 years of daily averaged velocity data is shown in the top panel. The second panel shows the probability of a front passing through each $0.1^\circ \times 0.1^\circ$ box in our domain over the model bathymetry. Small values have been made transparent. As a function of longitude, the third panel shows the separation between the mean latitude of the front and; i) the 1st and 99th percentile latitudes, ii) the 5th and 95th percentile latitudes, iii) the 25th and 75th percentile latitudes. This panel highlights regions along the front that exhibit large meanders. As a function of longitude, the fourth panel shows the separation between; i) the 1st and 99th percentile latitude of the front, ii) the 5th and 95th percentile latitude of the front, iii) the 25th and 75th percentile latitude of the front. Shown in red is the average total cross-front area flux as a function of longitude which highlights a correlation between regions of high cross-front transport and regions of frontal meandering.}
\label{fig:fronts_fig5}
\end{figure}

Simulations and observational studies have shown `hot spots' of enhanced eddy kinetic energy (EKE) and eddy-induced cross-jet transport can occur downstream of shallow bathymetric obstacles \cite{Chelton1990,Frenger2015,Patara2016,Barthel2022,FoppertDonohueWatts2017}. Hence, we expect to see enhanced eddy kinetic energy in locations of large total cross-front transport. We plot the EKE field at 70m depth computed over the 6 years of data (from model date 01/01/2008 to 30/12/2013) in the first panel of Figure \ref{fig:fronts_fig5}. Here, we see patches of increased EKE downstream of prominent bathymetry. Comparing these locations with the locations of large cross-front transport in Figures \ref{fig:fronts_fig3} and \ref{fig:fronts_fig4}, we find large cross-front transport typically occurs in, or downstream of, enhanced EKE.

In Figures \ref{fig:fronts_fig3} and \ref{fig:fronts_fig4} we observe a significant variation in the latitudinal range in the locations of cross-front transport, particularly where the maxima in cross-front transport occur. This variability results from meandering of the front due to interaction with bathymetry, leading to enhanced eddy kinetic energy and the observed increase in cross-front transport. The second panel of Figure \ref{fig:fronts_fig5} shows the likelihood of the front passing through any $0.1^\circ \times 0.1^\circ$ longitude-latitude box in our domain at any given time in our time series of front locations. It is effectively a histogram of the front locations normalized by the length of the observation period. This is superimposed on top of bathymetry. We observe a broader distribution of front location likelihoods in each of the regions that show particularly strong average southward and northward transport. A broader distribution of front location likelihood corresponds to a wider latitudinal range of front locations and identifies meandering of the front in that particular location. To visualize the locations of strong frontal meandering, and demonstrate how far from the mean the front will typically meander, we plot the separations between the mean and percentiles from a distribution of latitudes that the front passes through for each $0.1^\circ$ of longitude in the third panel of Figure \ref{fig:fronts_fig5}. To quantify the relationship between the latitudinal variability of the front, and the total cross-front transport, in the fourth panel of Figure \ref{fig:fronts_fig5} we plot the separations between pairs of percentiles from a distribution of latitudes that the front passes through for each $0.1^\circ$ of longitude.

These plots show an increased variability in the latitudinal location of the front, particularly in regions of strong cross-front transport. In fact, each series of latitudinal separations is strongly correlated to the average total cross-front area flux (shown as a red line in the fourth panel of Figure \ref{fig:fronts_fig5}). The Pearson correlation between the average total cross-front area flux and; (i) the 1 and 99th percentile series of latitudinal separations is $0.7144$; (ii) the 5 and 95th percentile series of latitudinal separations is $0.7066$; and (iii) the 25th and 75th percentile series of latitudinal separations is $0.6562$. These correlations suggest that the meandering experienced by the front is likely the dominant factor leading to increased cross-front transport along the front.

\subsection*{Choke points and meridional narrowing of the ACC}

We have shown that regions of strong frontal meandering are associated with increased cross-front transport. Conversely, cross-front transport is inhibited in regions where the fronts pass through bathymetric `choke points'. Specifically, a choke point can be defined as a region where neighboring fronts become narrowly spaced as they pass through a constriction, for example, between landmasses or sea floor obstacles. Choke points are associated with increased zonal flow, as streamlines likewise become narrowly spaced leading to a steepening in the meridional gradient of the stream function, which corresponds to an increase in the ACC flow speed. The fast flow through these choke points can inhibit meridional transport \cite{NaveiraGarabato2011}.

\cite{Park2019} characterize choke points by regions of narrowly concentrated frontal zones, which they define as contours of the Mean Dynamic Topography field, a field which represents the mean surface geostrophic streamlines. They identify five major choke points along the ACC: in the Udintsev fracture zone, the Drake passage, the Southwest Indian Ridge, Kerguelen plateau, and the Campbell plateau. In a analogous fashion, we identify choke points as regions where neighboring fronts (as identified by material curves rather than geostrophic streamlines) become closely spaced. Two additional sequences of fronts were constructed to track the northern and southern fronts identified in Figure \ref{fig:fronts_fig1} (labelled $\partial A_{3}$ and $\partial A_{1}$ respectively). We compute the meridional separation between these fronts as a function of longitude. As a simple hypothesis, we expect to see a narrowing of the fronts in regions of strong ACC flow speed, and conversely a widening of the fronts in regions of a weak ACC flow speed. However, deflection of fronts due to bathymetry and the interaction of the ACC with other current systems and features (such as the global gyres) may also play a role.

\begin{figure}[H]
\centering
\includegraphics[width=\linewidth]{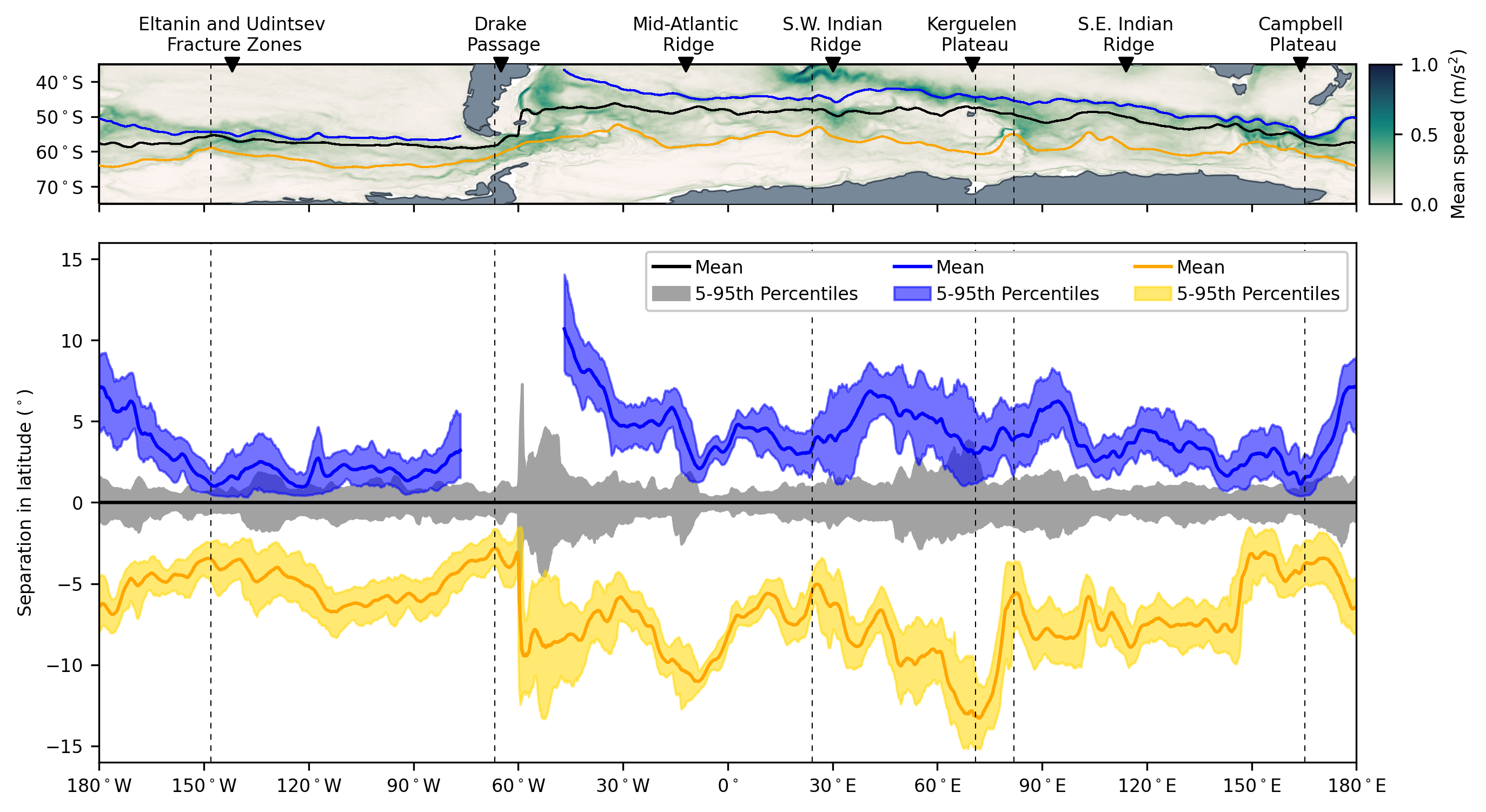}
\caption{\textbf{Antarctic Circumpolar Current choke points identified from three Southern Ocean fronts.} Major choke points in the ACC are identified as locations where the distance between Southern Ocean fronts decreases. Top panel: average latitude positions for three fronts (northern in blue, middle in black, southern in yellow). Vertical black dashed lines indicate the location of choke points where fronts become closely spaced. Two dashed lines are shown around the Kerguelen plateau to indicate where the northern/middle fronts and middle/southern fronts become closely spaced. These are superimposed on the average flow speed computed at 70m depth and show that choke points occur in regions where the average ACC flow speed increases. In most instances the northern front is disconnected at the Drake passage. Bottom panel: solid lines show average latitudinal separation between northern front (blue) and southern front (yellow) relative to the middle front (black). Shaded regions show 5th to 95th percentile range in front separations. We only plot data for the northern front for longitudes with an associated latitude value.
}
\label{fig:fronts_fig6}
\end{figure}

We plot the average latitude of the three fronts (northern $\partial A_{3}$ in blue, middle $\partial A_{2}$ in black, southern $\partial A_{1}$ in yellow) per $0.1^\circ$ of longitude for each of the three sequences of fronts in the top panel of Figure \ref{fig:fronts_fig6}, over the average flow speed at $70$m depth computed from 6 years of daily averaged velocity data. We identify five locations where the meridional separations between the three fronts diminish.
Moving from left to right in Figure \ref{fig:fronts_fig6}, the first choke point occurs in the Eltanin and Udintsev fracture zones, where the ACC speeds up as it is channeled between the fracture zones. The second occurs in the Drake passage, a well-known choke point of the ACC due to the South American and Antarctic land barriers to the north and south, and an increase in the average ACC flow speed. The third occurs over the Southwest Indian ridge, where we see a noticeable increase in the ACC flow speed along the middle front. The fourth occurs on the Kerguelen plateau. Here we see the southern front is steered southward around the plateau, as the middle and northern fronts narrow, until the southern front diverts northward again closing the gap to the middle front. The fifth ACC choke point occurs near two significant bathymetric obstacles. The three fronts narrow as the ACC passes over the South East Indian ridge, south of Tasmania, and remain narrowly spaced as they are steered around the Campbell plateau. This region exhibits a significantly higher average ACC flow speed compared to the rest of the ACC. These five choke points are consistent with those identified by \cite{Park2019}, and occur in regions of relatively strong ACC flow speeds, where the ACC is steered by bathymetric obstacles. The fronts tend to diverge downstream of each of these five choke points, which correspond to the regions of increased meandering and cross-front transport we observed along the middle front in Figure \ref{fig:fronts_fig5}.

To shed further light on these choke points, we compute percentiles of the latitudinal separations between pairs of fronts. Specifically, in the bottom panel of Figure \ref{fig:fronts_fig6}, solid lines show average latitudinal separation between northern front (blue) and southern front (yellow) relative to the middle front (black). Shaded regions show 5th to 95th percentile range in front separations. Also shown is 5th to 95th percentile range of the middle sequence of fronts (shaded gray, as shown in Figure \ref{fig:fronts_fig5}).
These statistics confirm that the locations of the five identified choke points are consistent over the 4.5-year sequences of fronts.

\section{Discussion}

In this work we have investigated Southern Ocean fronts and cross-front transport using six years of daily particle trajectories advected with velocities at 70m depth from a $0.1^\circ$-resolution global ocean model (ACCESS-OM2; \cite{Kiss2020}). Our approach uses the theory of finite-time coherent sets, which identify regions of phase space that minimally mix. The boundary of a coherent set defines a material curve that represents the location of a Southern Ocean front. These fronts act as transport barriers between regions of the ocean with different water properties (see Figure \ref{fig:fronts_fig1}), and best resist distortion and filamentation over a finite-time window (see Figure \ref{fig:fronts_ftcs_diagram}). By comparing fronts from a sequence of sliding windows and identifying regions of ocean that enter or exit the front, we located and quantified northward and southward cross-front transport (see Figure \ref{fig:fronts_fig2}). Our results reveal a previously unobserved global pattern of alternating poleward and equatorward cross-front transport (see Figure \ref{fig:fronts_fig3}).

We found that large cross-front transport occurs in regions downstream of seven major bathymetric features (see Figure \ref{fig:fronts_fig4}): the Eltanin and Udintsev fracture zones, the Drake passage, the Mid-Atlantic ridge, the Kerguelen plateau, the Mid-Atlantic ridge, the Southeast Indian ridge, and the Campbell plateau. Our identified fronts tend to meander downstream of these bathymetric features, inducing stronger cross-front transport. This coincides with locations of and downstream of increased eddy kinetic energy (see Figure \ref{fig:fronts_fig5}). The locations of increased cross-front transport are consistent with previous studies \cite{Dufour2015, Thompson2012}, and highlight the important role bathymetry plays in influencing the locations of increased cross-front transport. In contrast to \cite{Dufour2015, Thompson2012}, however, our results identify a distinct geographic pattern of alternating northward and southward cross-front transport. Furthermore, by considering two additional sequences of front locations, one to the north and one to the south, we identified five choke points along the ACC, consistent with \cite{Park2019} who used traditional (non-material) front definitions and observational datasets. Our identified choke points occur near major bathymetric obstacles, where the ACC and fronts tend to narrow as the ACC flow is steered by the bathymetry in order to conserve potential vorticity (see Figure \ref{fig:fronts_fig6}). The locations of these choke points are important, as they identify regions where the ACC flow is constrained by bathymetric obstacles, leading to an increase in the average ACC flow speed which inhibits cross-front transport.

The method used in this study relies on the kinematics of the Southern Ocean alone, requiring no hydrographic data. We use particle trajectories advected using a high-resolution ocean model to construct each sequence of front locations. These virtual particles are fixed at 70m depth, ignoring vertical velocities, in order to reduce the effects of the Ekman layer. This work represents the first mixed boundary condition approach to identify finite-time coherent sets and transport barriers in an oceanographic setting. We note that our approach can be adapted to multiple dimensions, for example, one could define a Southern Ocean front as a surface (rather than a curve) between minimally mixing water volumes. In this framework, one could estimate cross-front volume transport (in Sverdrups), and quantify tracer fluxes across the front through time (e.g., heat flux, salt flux, and carbon flux); this may shed light on the important mechanisms underpinning cross-front transport in the true three-dimensional context. Our approach and results also motivate further study of Southern Ocean fronts from a Lagrangian perspective, particularly as more surface drifters and ARGO floats are deployed in the Southern Ocean in future years, and simulations of the complex Southern Ocean dynamics are resolved at ever higher resolutions.

\section{Methods}
\subsection*{Theoretical constructions}
\subsubsection*{Finite-time coherent sets with mixed boundary conditions}
We choose a domain $\mathcal{M}$ that covers the region of ocean south of $30^\circ$S at 70m depth with land removed.
Denote by $\partial \mathcal{M}^D$ the boundary of $\mathcal{M}$ corresponding to the Antarctic coastline, where we impose Dirichlet boundary conditions.
Denote by $\partial \mathcal{M}^{N}$ all non-Antarctic boundaries (which encompasses both non-Antarctic coastlines and open water boundaries), where we impose Neumann boundary conditions.
For $t\in \mathbb{R}$ we consider smooth volume-preserving flow maps $\Phi^{t}$ that describe the evolution of points $x\in \mathcal{M}$ to their future locations $\Phi^{t}(x)$ at time $t$, with inverse flow maps $(\Phi^{t})^{-1}$ describing the reverse evolution, which we denote by $\Phi^{-t}$.
Denote by $\text{int}(\mathcal{M})$ the interior of $\mathcal{M}$, which is $\mathcal{M}$ with its boundary subtracted. Let $A\subset \text{int}(\mathcal{M})$ be a set that completely surrounds Antarctica, for example, the dark gray regions shown in Figure \ref{fig:fronts_fig2}.
We seek the family of evolving sets $\Phi^{t}(A)$ that remains minimally distorted over the time window $W(0,T) = [-T/2, T/2]$ of length $T>0$ centered at time $t_c=0$. In this methods section we have nominally chosen $t_c=0$ for brevity. Here, finite-time coherent sets are regions $A$ that yield close to minimal values of the dynamic Cheeger constant \cite{Froyland2018}, which quantifies the average evolved boundary length relative to enclosed area:

\begin{equation} \label{eq:dynamic_cheeger}
    \mathbf{h}^{D}(A) := \frac{\frac{1}{T}\int_{-T/2}^{T/2} \text{length}(\Phi^{t}(\partial A)) \text{ d}t} {\text{area}(A)}.
\end{equation}

In this framework, we define \emph{circumpolar} Southern Ocean fronts as the northern boundary of a finite-time coherent set $A$ whose boundary $\partial A$ surrounds Antarctica.
Boundaries of finite-time coherent sets, and therefore fronts, can be approximated by a level set of the dominant eigenfunction of a \emph{dynamic Laplace operator} \cite{Froyland2015a}, which we now define.
The connection between the dynamic Laplace operator and Eq. (\ref{eq:dynamic_cheeger}) is developed in \cite{Froyland2018}.
The natural forward-in-time action of $\Phi^t$ on scalar-valued functions $f$ is $\Phi^{t}_{*} f := f \circ \Phi^{-t}$, where $\Phi^{t}_{*}$ is known as a push-forward.
Similarly, the natural backward-in-time action on scalar-valued functions is the pull-back $(\Phi^{t})^{*}$
defined by $(\Phi^{t})^{*}f = f \circ \Phi^{t}$.
The push-forward and pull-back operators allow us to evolve smooth functions $f$ anchored at the center time $t_c=0$ on $\mathcal{M}$.
Denoting by $\Delta$ the usual Laplace operator, we now define the dynamic Laplace operator over the window $W(0,T)$ as
\begin{equation}\label{eq:dynamic_laplace}
\Delta^{D}_{W(0,T)}f := \frac{1}{T}\int_{-T/2}^{T/2} (\Phi^{t})^{*}(\Delta(\Phi^{t}_{*}f))\ dt,
\end{equation}
with associated mixed boundary condition eigenproblem
\begin{align*}\label{eq:mixed_eigenproblem}
\Delta^D_{W(0,T)} v &= \lambda v \quad \text{ on int}(\mathcal{M}),\\
\int_{-T/2}^{T/2} \frac{\partial \Phi^{t}_{*} v}{\partial n^{t}}\circ \Phi^{t}\text{ d}t &= 0 \quad \text{ on } \partial \mathcal{M}^{N},\\
v &= 0 \quad \text{ on } \partial \mathcal{M}^{D},
\end{align*}
where $n^{t}$ is the outward normal to $\partial \mathcal{M}^{N}$. In practice, we form and solve the following weak form of the eigenproblem on int($\mathcal{M}$), and apply Dirichlet boundary conditions $v=0$ on $\partial \mathcal{M}^{D}$,
\begin{equation*}\label{eq:weak_igenproblem}
-\frac{1}{T}\int_{-T/2}^{T/2}\int_{\Phi^{t}(\mathcal{M})} \nabla(\Phi^{t}_{*} v) \cdot \nabla(\Phi^{t}_{*}\psi)\ d\ell = \lambda \int_{\mathcal{M}} v\psi\ d\ell \quad \text{for all }\psi \in H^{1}(\mathcal{M}) \text{ with } \psi=0\mbox{ on $\partial \mathcal{M}^{D}$}.
\end{equation*}
By solving this weak form of the eigenproblem, we automatically satisfy the Neumann boundary conditions on $\partial \mathcal{M}^{N}$.

Due to the approximately circular motion of the ACC around Antarctica, we expect the level contours of the leading eigenfunction $v$ of $\Delta^{D}_{W(0,T)}$ to provide a family of fronts, parameterized by the level value.
These level contours represent a front anchored at time $t_c=0$.
More precisely, the level contour of $v$ at level value $a$ is denoted $\partial A_a := \{x \in \mathcal{M}\ :\ v(x) = a \}$. For example, in Figure \ref{fig:fronts_fig1}, three fronts are plotted from south to north; $\partial A_{a_1}$, $\partial A_{a_2}$, and $\partial A_{a_3}$, where $a_1=0.55$, $a_2=0.638$, and $a_3=0.6775$. (Note that, for brevity, we referred to these level contours as $\partial A_1$, $\partial A_2$, $\partial A_3$ in the main text.) These fronts are level sets of $\Delta^{D}_{W(270,540)}$, anchored at time $t_c=270$, and correspond to the northern boundaries of finite-time coherent sets defined by sublevel sets of $v$. Specifically, $A_a := \{x \in \mathcal{M}\ :\ v(x) \le a \}$. The southern boundary of $A_a$, the Antarctic coastline, is simply the zero level set (where $a=0$).
To illustrate, a sketch of a finite-time coherent set $A$ is shown as the dark gray region in the top left of Figure \ref{fig:fronts_fig2}), where the front $\partial A$ is shown as black boundary.

\subsubsection*{Tracking sequences of finite-time coherent sets and fronts from sliding windows} \label{sec:tracking}
So far we have described an approach to identify a front, anchored at time $t_c=0$, which depends on the choice of the time window $W(0,T)$. We will now vary the window length $T$ and/or the center time $t_c$, leading to a two-parameter family of windows $W(t_c,T) = [t_c-T/2,t_c+T/2]$. Differentiable changes to the window are expected to lead to differentiable changes to the coherent set \cite{AFJ21}. We can construct corresponding \emph{sequences of coherent sets} by systematically varying one or both of these parameters. As defined in \cite{Denes2022}, varying $T$ results in a \textit{telescoping sequence}, while varying $t_c$ results in a \textit{sliding sequence}. For each window centered at time $t_c$, we will define the domain $\mathcal{M}_{t_c}$ as the region of global ocean south of $30^\circ$S, at 70m depth, at time $t_c$.

Both telescoping and sliding sequences of coherent sets can reveal important characteristics of the coherent object being studied. Recent studies have used telescoping windows to investigate their effect on the geometry of finite-time coherent sets \cite{Froyland2014a}, to determine the life expectancy of an ocean eddy \cite{ AndradeCanto2020}, and to find a maximal coherence timescale of a finite-time coherent which minimizes the rate of mass loss per unit window length through the boundary of the coherent set \cite{Denes2022}. A sliding sequence of windows have been used to study the decay of an ocean eddy \cite{Froyland2015c, Denes2022}, to investigate the Lagrangian pathways of convective heat transfer in turbulent Rayleigh-B\'enard convection flows, and to find structural changes in coherent features (such as merging or splitting) \cite{blachut2020, blachut2021, ndour2021}. An alternative approach used to determine when a coherent set is born or dies was developed in \cite{Froyland2021}, which removes the need for multiple sliding and telescoping computations by considering an inflated dynamic Laplace operator on a time-expanded domain.

In our oceanographic context, we use a sliding sequence of coherent sets to quantify transport across circumpolar Antarctic ocean fronts. To do this, we match a particular front anchored at time $t_c$ to the previous front anchored at $t_c-1$ , or the subsequent front anchored at $t_c+1$. The flow map from time $t_c$ to $t_c+t$ will be denoted as $\Phi_{t_c}^{t}=\Phi^{t_c+t}_0\circ (\Phi^{t_c}_0)^{-1}$. For clarity, we denote by $v_{t_c}$ the dominant eigenfunction of $\Delta^{D}_{W(t_c,T)}$ anchored at time $t_c$.

To `match' fronts from one center time to the next we will use an `equal-area’ constraint. Let $\{t_{\min},\dots,t_{\max}\}$ be a sequence of center times over which we wish to track a front, spaced 1 unit of time apart. We start by constructing a sequence of windows $\{W(t_c,T)\}_{t_c\in \{t_{\min}, \dots, t_{\max}\}}$. We identify a particular level set value $a(t_{\min})$ of the eigenfunction $v_{t_{\min}}$ corresponding to a particular front of interest $\partial A_{a(t_{\min})}$ anchored at time $t_{\min}$. Denote by $\mathcal{A}:= \text{area}\left(A_{a(t_{\min})}\right)$ the area contained south of the front $\partial A_{a(t_{\min})}$, or simply the area of the corresponding finite-time coherent set $A_{a(t_{\min})}$.
We identify the location of all subsequent fronts at times $t_c>t_{\min}$ (anchored at time $t_c$) by finding the level set value $a(t_c)$ such that the area of the finite-time coherent set $A_{a(t_c)}$ is equal to $\mathcal{A}$.

\subsubsection*{Locating and quantifying cross-front transport} \label{sec:materialexchange}

We now describe an approach to locate and quantify cross-front transport. We begin with a sliding sequence of front locations $\{\partial A_{a(t_c)} \}_{t_c\in \{t_{\min}, \dots, t_{\max}\}}$, identified in the previous subsection. Particles may move back and forth across the front due to daily fluctuations in the front position. To minimize this, we will identify locations and quantities of cross-front transport between two fronts anchored $s$ days apart (in the Results section we use $s=30$ days). Because the fronts we identify are predominantly zonal we will speak of `entering' transport as southward transport, and `exiting' transport as northward transport. For a particular front $\partial A_{a(t)}$, we want to identify regions of fluid that will enter southwards across the sequence of fronts in $s$ days, as well as the regions of fluid that will exit northwards across the sequence of fronts in $s$ days. To do this, we compare the region contained in the finite-time coherent set $A_{a(t)}$ anchored at time $t$ with the region contained in the $s$-day pull-back of the finite-time coherent set $A_{a(t+s)}$ anchored at time $t+s$. More precisely, we compare $A_{a(t)}$ and $\Phi^{-s}_{t+s}\left(A_{a(t+s)}\right)$.
Denote by $\mathcal{E}_{t}$ the region of ocean that `enters' $\Phi^{-s}_{t+s}\left(A_{a(t+s)}\right)$, i.e. is not contained in $A_{a(t)}$ but is contained in $\Phi^{-s}_{t+s}\left(A_{a(t+s)}\right)$, and denote by $\mathcal{X}_{t}$ the region of ocean that `exits' $A_{a(t)}$, i.e. is contained in $A_{a(t)}$ but is not contained in $\Phi^{-s}_{t+s}\left(A_{a(t+s)}\right)$;  see Figure \ref{fig:fronts_fig2}. More formally,

\begin{equation}
    \mathcal{E}_{t} := \Phi^{-s}_{t+s}\left(A_{a(t+s)}\right) \backslash A_{a(t)}, \text{ and } \mathcal{X}_{t} :=  A_{a(t)} \backslash \Phi^{-s}_{t+s}\left(A_{a(t+s)}\right).
\end{equation}

Since our sliding sequence of fronts is defined using an equal-area constraint, the total southward cross-front transport (entry regions $\mathcal{E}_t$) must be offset by the total northward transport (exit regions $\mathcal{X}_t$) along the front. That is, $\text{area}(\mathcal{E}_{t}) = \text{area}(\mathcal{X}_{t})$, which implies that along the entire front $\partial A_{a(t)}$, there is net zero cross-front transport. This is the simplest and most physically reasonable constraint we could impose in order to track a front from one window to the next. We construct a time series of northward and southward cross-front transport regions ($\{\mathcal{X}_t\}_{t\in\{t_{\min}, \dots, t_{\max}-s\}}$ and $\{\mathcal{E}_t\}_{t\in\{t_{\min}, \dots, t_{\max}-s\}}$ respectively), where this net-zero transport property holds for all center times $t_c \in \{t_{\min}, \dots, t_{\max}-s\}$.
Taking a temporal average of this time series over each point in space, we can determine whether a front exhibits a typical pattern of cross-front transport, and quantify the average amount of northward, southward, and net cross-front transport for each location. More formally,
\begin{equation}\label{eq:avg_flux}
    \overline{\mathcal{E}} := \frac{1}{(t_{\max}-s)-t_{\min} + 1} \sum_{t=t_{\min}}^{t_{\max}-s} \mathds{1}_{\mathcal{E}_t}, \text{ and } \overline{\mathcal{X}} := \frac{1}{(t_{\max}-s)-t_{\min} + 1} \sum_{t=t_{\min}}^{t_{\max}-s} \mathds{1}_{\mathcal{X}_t},
\end{equation}
where $\overline{\mathcal{E}}(x)$ (respectively, $\overline{\mathcal{X}}(x)$) is the time-averaged southward (respectively, northward) transport at position $x$, and $\mathds{1}_{\mathcal{E}_t}$ is the indicator function defined on $\mathcal{M}_t$. That is, $\mathds{1}_{\mathcal{E}_t}(x) = 1$ for a point $x\in \mathcal{E}_t$, and $\mathds{1}_{\mathcal{E}_t} = 0$ elsewhere. This process is sketched in Figure \ref{fig:fronts_fig2}.

\subsection*{Numerical constructions}

\subsubsection*{Ocean model and trajectory integration}\label{sec:model}
The only data required in our approach are particle trajectories, which are used to approximate flow maps $\Phi^{t}$. To construct a set of particle trajectories we use daily averaged velocity data derived from the Australian Community Climate and Earth System Simulator global ocean-sea ice model (ACCESS-OM2) \cite{Kiss2020}. The model couples the MOM5.1 ocean model to the CICE5.1.2 sea ice model via OASIS3-MCT and is forced using prescribed atmospheric conditions from the JRA55-do reanalysis. We use the ACCESS-OM2-01 `eddy-rich' configuration of the model, which has a horizontal resolution of $0.1^\circ$ between $65^{\circ}$S and $65^{\circ}$N, and 75 vertical levels, and fully resolves the active mesoscale eddy field in most of the ocean \cite{Kiss2020}. A particular motivation for the development of this model was to better represent the dynamics of the Southern Ocean, where mesoscale variability plays a crucial dynamical role in the system's evolution \cite{Hogg2015}. We used the conservative temperature, practical salinity, and effective sea level data from this model in Figure \ref{fig:fronts_fig1}, and in the animation and Figures \ref{fig:supp_fronts_temp},\ref{fig:supp_fronts_salt},\ref{fig:supp_fronts_ssh} in the Supplementary Material.

We initialize $60,000$ particles fixed at $70$m depth in the Southern hemisphere, $D=[180^\circ$W$,180^\circ$E$)\times [90^\circ$S$,0^\circ]$, on a Fibonacci lattice \cite{Gonzalez2009} which approximately equidistribute the particles over the domain. We restrict these particles to $70$m depth in order to reduce the wind-induced transport effects within the Ekman layer. Particles are advected forward in time using daily averaged $70$m horizontal velocity fields (neglecting vertical velocities) with the Lagrangian ocean analysis tool \texttt{Parcels} \cite{Delandmeter2019, Lange2017} from the model date 01/01/2008 for $2190$ days with a $5$-minute timestep using the fourth-order Runge-Kutta advection scheme with partial-slip boundary conditions. This small timestep is used in order to reduce possible particle overshoot and potential errors due to the advection scheme. We remove approximately $13,000$ particles that were initialized on land, resulting in a total of $n = 47,393$ valid particle trajectories. In the case where particles become `stuck' on coastlines, we treat those particles as `missing' from the date they become stuck.

The particle positions $x^{t}_{i} \in D$ are stored as longitude-latitude coordinates $(\alpha^{t}_{i}, \beta^{t}_{i})$ where $i \in \{1, \dots, n\}$ is the particle index and $t \in \tau = \{0,1,\dots,2190\}$ are observation times measured in days from release. The trajectories are used to approximate flow maps in this region over the time interval $[0,2190]$. We denote by $\Phi^{t}_{0}$ the flow map generated by particles released on day $0$ and advected for $t$ days, i.e., $x^{t}_{i} = \Phi^{t}_{0}(x^{0}_{i})$, where $x^{0}_{i}$ is the initial position of the $i$th particle.

\subsubsection*{Domain selection and meshing} \label{sec:mesh}
In order to focus on transport barriers in the Southern Ocean, we restrict our domain to the region of ocean south of $30^\circ$S. Specifically, for each window $W(t_c,T)$, we define the domain $\mathcal{M}_{t_c} := \{x^{t_c}= (\alpha^{t_c}, \beta^{t_c}) \in \Phi_{0}^{t_c}(\mathcal{M})\ :\ \beta^{t_c} \leq 30^\circ\text{S} \}$.  The indices of particles contained in $\mathcal{M}_{t_c}$ are simply $I_{t_c} := \{i : \beta_{i}^{t_c} \leq 30^\circ\text{S} \}$.
Additionally, we define a discrete set of $5,128$ coastline points $C = \{c_j\}_{j \in \{1,\dots,5128\}}$ obtained from the Natural Earth coastline dataset at the $1:110$ million scale included in the Python package \texttt{cartopy}. We distinguish between coastline points along the Antarctic coast as $C^D$ (where we apply Dirichlet boundary conditions), and all other non-Antarctic coastline points as $C^N$, where $C^D \cap C^N = \emptyset$ and $C^D \cup C^N = C$. Here $|C^D| = 650$ and $|C^N| = 4478$.

The finite-element method we use to numerically approximate the dynamic Laplace operator requires a triangular mesh of $\mathcal{M}_{t_c}$, as well as the meshes of $\Phi^{t-t_c}_{t_c}(\mathcal{M}_{t_c})$ for $t \in W(t_c,T)$. To do this we modify Algorithm 1 developed in \cite{Denes2022}. This algorithm takes input of: a chosen time $t \in W(t_c,T)$, the particle set $\{x^{t}_{i}\}_{i= 1,\dots,n}$, and indices $I_{t_c}$. The output is a mesh $(V,E)$ and the boundary edges $\partial E$ of $\Phi^{t-t_c}_{t_c}(\mathcal{M}_{t_c})$. We make the following modifications. We identify and remove `stuck' particles from the particle set (particles whose positions do not change in the subsequent timestep). The input particle set is now $\{x_{i}^{t}\}_{i\in I_{t_c}} \cup C^A \cup C^N$ (with `stuck' particles removed) rather than the entire particle set. This replaces the need for the input of indices $I_{t_c}$. Since we apply Dirichlet boundary conditions to the Antarctic coastline, triangles that cover Antarctic land have no impact on the eigenfunctions of the dynamic Laplacian as the eigenfunction is set to zero at their vertices. The only land triangles we are required to remove are triangles that cover non-Antarctic land. To remove these non-Antarctic land triangles, in Step 3, we let the set of faces whose vertices all lie in $\Phi^{t-t_c}_{t_c}(\mathcal{M}_{t_c})$ to be $\mathcal{F}' = \{\mathcal{F}(f) \in \mathcal{F} :\ \mathcal{F}(f)\in I_{t_c}\cup\{n+1,\dots,n+|C^{D}|\}\}$. That is, we keep triangles within the triangulation whose vertices are all in the $I_{t_c}$ or correspond to Antarctic coastline points. This removes all non-Antarctic coastline points, all triangles with at least one vertex that is a non-Antarctic coastline point, and hence all non-Antarctic land triangles from the triangulation. Alternate approaches to removing land triangles exist, however, our approach retains all regions of ocean around Antarctica that are poorly sampled by particle trajectories while being very quick to construct in practice. An illustration of such a mesh is shown in Figure \ref{fig:mesh} in the Supplementary Material (where we have removed Antarctic land triangles).

\subsubsection*{Numerical method to compute eigenfunctions of the dynamic Laplacian} \label{sec:data_numerics}
A range of numerical methods exist to approximate the eigenvalues and eigenfunctions of the dynamic Laplace operator \cite{FJ2015,Froyland2015a,Froyland2018}. We use the adaptive transfer operator method developed in \cite{Froyland2018, Denes2022} based on the finite-element method. Specifically, we follow Sections III, III (C), and III (D) of \cite{Denes2022}. To deal with `stuck' particles, we follow Section 3.2.4 of \cite{Froyland2018}.

\subsubsection*{Computing the location of a front}
To compute the front $\partial A_{a(t_c)}$, working in a South Polar Stereographic coordinate system, we use the \texttt{tricontour} function of the Matplotlib package in python, which takes input of the eigenfunction $v_{t_c}$, the mesh of $\mathcal{M}_{t_c}$, and the level set value $a(t_c)$. The output of the \texttt{tricontour} function is a list of points $z_{1},\dots, z_{m}$ which form a discrete representation of the front.

\subsubsection*{Tracking sequences of finite-time coherent sets and fronts from sliding windows} \label{sec:tracking_numerics}
We first identify a front of interest at time $t_{\min}$, which determines the level set value $a(t_{\min})$. We approximate the area $\mathcal{A}:= \text{area}\left(A_{a(t_{\min})}\right)$ by distributing approximately 256 million points over the Earth's surface on a Fibonacci lattice \cite{Gonzalez2009}. By construction, each point on the lattice covers a surface area of 1 km$^2$. Denote by $Q$ this set of equidistributed points. Working in a South Polar Stereographic coordinate system, we use the \texttt{LinearTriInterpolator} function from the Matplotlib package in python to linearly interpolate the eigenfunction $v_{t_{\min}}$ using the mesh of $\mathcal{M}_{t_{\min}}$ onto the points in $Q$, resulting in a 256 million $\times\ 1$ array of interpolated values. Since our mesh contains triangles across Antarctic land, the points in $Q$ contained inside these Antarctic land triangles will have an interpolated value of zero, while points in $Q$ that cover non-Antarctic land (and are outside of the mesh of $\mathcal{M}_{t_{\min}}$) will have `Not a Number' interpolated values. The area of $A_{a(t_{\min})}$, measured in km$^2$, is then approximated by counting the number of entries in the array of interpolated values that are less than or equal to $a(t_{\min})$ (we include Antarctic land).

With $\mathcal{A}$ calculated, we next find $a(t_c)$ for $t_c>t_{\min}$ by linearly interpolating the eigenfunction $v_{t_c}$ using the mesh of $\mathcal{M}_{t_c}$ onto the points in $Q$, resulting in a new 256 million $\times\ 1$ array of interpolated values. To find $a(t_c)$, we sort the array of interpolated values in ascending order and identify the interpolated value in the $\mathcal{A}$th entry, which corresponds to $\text{area}\left(A_{a(t_c)}\right)$. In practice, to identify the $\mathcal{A}$th entry we use the \texttt{partition} function from the numpy package in python. We repeat this process to construct a sequence of level set values $\{a(t_c)\}_{t_c\in\{t_{\min},\dots,t_{\max}\}}$.

Since the dynamics of $\Phi^{t}$ are not entirely divergence-free, there are small differences between the areas of $A_{a(t_c)}$ and the $s$-day pull-back $\Phi^{-s}_{t_c}\left(A_{a(t_c)}\right)$. To ensure $A_{a(t_c)}$ and $\Phi^{-s}_{t_c}\left(A_{a(t_c)}\right)$ have equal area, we construct an additional sequence of level set values, \\
$\{a'(t_c)\}_{t_c\in\{t_{\min},\dots,t_{\max}\}}$, such that $\text{area}\left(\Phi^{-s}_{t_c}\left(A_{a'(t_c)}\right)\right) = \mathcal{A}$ for all $t_c \in \{t_{\min},\dots,t_{\max}\}$.

\subsubsection*{Locating and quantifying cross-front transport} \label{sec:materialexchange_numerics}
We start by distributing approximately 256 million points over the Earth's surface on a Fibonacci lattice \cite{Gonzalez2009}, denoted by $Q$. To numerically approximate $\mathcal{E}_{t_c}$ and $\mathcal{X}_{t_c}$, we:
\begin{enumerate}
    \item interpolate the eigenfunction $v_{t_c}$ using the mesh of $\mathcal{M}_{t_c}$ onto the points in $Q$, resulting in a 256 million vector of interpolated values.
\item interpolate the eigenfunction $(\Phi^s_{t_c+s})^{*}v_{t_c+s}$ using the mesh $\Phi^{-s}_{t_c+s}(\mathcal{M}_{t_c+s})$ onto the points in $Q$ resulting in a second 256 million vector of interpolated values.
\end{enumerate}
We compare the above two vectors of interpolated values to identify points in $Q$ that are inside and/or outside $A_{a(t_c)}$ and $\Phi^{-s}_{t_c+s}\left(A_{a'(t_c+s)}\right)$, resulting in two lists of points contained in $\mathcal{E}_{t_c}$ and $\mathcal{X}_{t_c}$.

To represent these regions on a uniform grid, we start by constructing a $0.1^\circ \times 0.1^\circ$ uniform longitude-latitude grid over the domain $D$, which is represented by a $3600\times900$ array. In each grid box we count the number of points contained inside $\mathcal{E}_{t_c}$ or $\mathcal{X}_{t_c}$ resulting in two $3600\times900$ arrays, denoted by $\frak{E}_{t_c}$ and $\frak{X}_{t_c}$ respectively, that represent a southward and northward area flow rate in units of km$^2$ per $s$ days. We set $\frak{E}_{t_c}= \frak{E}_{t_c}/s$ and $\frak{X}_{t_c}= \frak{X}_{t_c}/s$, and divide each entry by the longitudinal extent of its corresponding box in km (accounting for latitude, where we use the southern edge of each box). After this division, $\frak{E}_{t_c}$ and $\frak{X}_{t_c}$ represent a southward and northward area flux in units of km/day. The southward area flux in the $(i,j)$ box is denoted by $\mathfrak{E}_{t_c}^{i,j}$.

To construct $\overline{\mathcal{E}}$ and $\overline{\mathcal{X}}$, visualized by the global patterns of cross-front transport in Figure \ref{fig:fronts_fig3}, we repeat the above process for all times $t_c \in \{t_{\min},\dots, t_{\max}\}$. This produces two time-indexed sets of arrays, $\{\frak{E}_{t_c}\}_{t_c = t_{\min}, \dots, t_{\max}-s}$ and $\{\frak{X}_{t_c}\}_{t_c = t_{\min}, \dots, t_{\max}-s}$. Finally, we compute the time-averaged southward and northward area fluxes as
\begin{equation}
\overline{\frak{E}}^{i,j} := \frac{1}{t_{\max}-s -t_{\min}+1} \sum_{t_c = t_{\min}}^{t_{\max}-s} \frak{E}_{t_c}^{i,j}, \text{ and } \overline{\frak{X}}^{i,j} := \frac{1}{t_{\max}-s -t_{\min}+1} \sum_{t_c = t_{\min}}^{t_{\max}-s} \frak{X}_{t_c}^{i,j}.
\end{equation}

To construct the area flux per $0.1^\circ$ longitude shown in Figure \ref{fig:fronts_fig4}, we compute $\sum_{i=1}^{900}\overline{\frak{E}}^{i,j}$ and $\sum_{i=1}^{900}\overline{\frak{X}}^{i,j}$, and multiply the resulting $3600 \times 1$ arrays by the latitudinal extent of each box in km ($0.1^\circ$ which is approximately 11.13 km$^2$).

\subsubsection*{Computing a front likelihood map} \label{sec:likelihood}

To numerically construct the front likelihood map shown (with resolution $0.1^\circ \times 0.1^\circ$) in Figure \ref{fig:fronts_fig5}, we start by constructing a $0.05^\circ \times 0.05^\circ$ uniform longitude-latitude grid over the domain $D$, represented by a $7200 \times 1800$ array, denoted by $D_1$. We use a grid spacing of $0.05^\circ$ to ensure we sample the front at least once per $0.1^\circ$ of longitude. At center time $t_c \in \{t_{\min},\dots, t_{\max}\}$, we linearly interpolate the eigenfunction $v_{t_c}$ using the mesh $\mathcal{M}_{t_c}$ onto the grid corners of $D_1$. The output of this interpolation is a $7200 \times 1800$ array, denoted by $\vartheta_{t_c}$. To approximate the front $\partial A_{a(t_c)}$, we apply the \texttt{tricontour} function of the Matplotlib package in python. This function takes as input the array $\vartheta_{t_c}$, the grid $D_1$, and the level set value $a(t_c)$. The output is a collection of points $z_1, \dots, z_{m'}$ which represent points along the front.

Due to the $0.05^\circ$ grid spacing, the separation between consecutive points $z_i$ and $z_{i+1}$ is at most $0.1^\circ$. To identify the latitudinal position/s of the front per $0.1^\circ$ of longitude, we construct a $0.1^\circ \times 0.1^\circ$ uniform longitude-latitude grid over the domain $D$, represented by a $3600 \times 900$ array denoted by $D_2$. We then identify the boxes in $D_2$ that contain at least one point $z_i \in \{z_{1},\dots, z_{m'}\}$. This results in a $3600 \times 900$ Boolean array denoted by $\gamma_{t_c}$, representing the boxes that the front passes through at center time $t_c$.

We repeat this process for all center times $t_c \in \{t_{\min}, \dots, t_{\max}\}$ to construct a time-indexed set of Boolean arrays $\Upsilon := \{\gamma_{t_c}\}_{t_c\in\{t_{\min},\dots,t_{\max}\}}$. We then compute the front likelihood map shown in Figure \ref{fig:fronts_fig5} by taking a time-average over $\Upsilon$. We also use $\Upsilon$ to compute the percentiles of separations shown in Figures \ref{fig:fronts_fig5} and \ref{fig:fronts_fig6}.

\section*{Data availability}
The ACCESS-OM2 data used in this study is hosted in the COSIMA Model Output Collection, doi: 10.4225/41/5a2dc8543105a. The processed data are available from the corresponding author on reasonable request.

\section*{Code availability}
The Parcels python code used to advect virtual particles is available at \url{www.github.com/OceanParcels/parcels}. Python code implementing the numerical techniques is available at \url{www.github.com/michaeldenes/FEMDLpy}.

\section*{Acknowledgments}
The authors would like to thank the COSIMA consortium (\url{www.cosima.org.au}) for making available the ACCESS-OM2 suite of models (see \url{www.github.com/COSIMA/access-om2}). This research was undertaken with the assistance of resources and services from the National Computational Infrastructure (NCI), which is supported by the Australian Commonwealth Government. M.C.D. is supported by an Australian Government Research Training Program Scholarship, G.F. is partially supported by an ARC Discovery Project, and S.R.K. is supported by the Australian Research Council Special Research Initiative, Australian Centre for Excellence in Antarctic Science (ARC Project Number SR200100008).

\section*{Author contributions}
All authors contributed to the all aspects of the study. Computations were performed by M.C.D. M.C.D wrote the first draft of the manuscript, and all authors contributed to editing the manuscript.

\section*{Competing interests}
The authors declare that they have no competing interests.

\bibliography{library}
\bibliographystyle{siam}

\pagebreak

\setcounter{equation}{0}
\setcounter{figure}{0}
\setcounter{table}{0}

\renewcommand{\theequation}{S\arabic{equation}}
\renewcommand{\thefigure}{S\arabic{figure}}

\section{Supplementary material}

\begin{figure}[H]
\centering \includegraphics[width=0.8\linewidth]{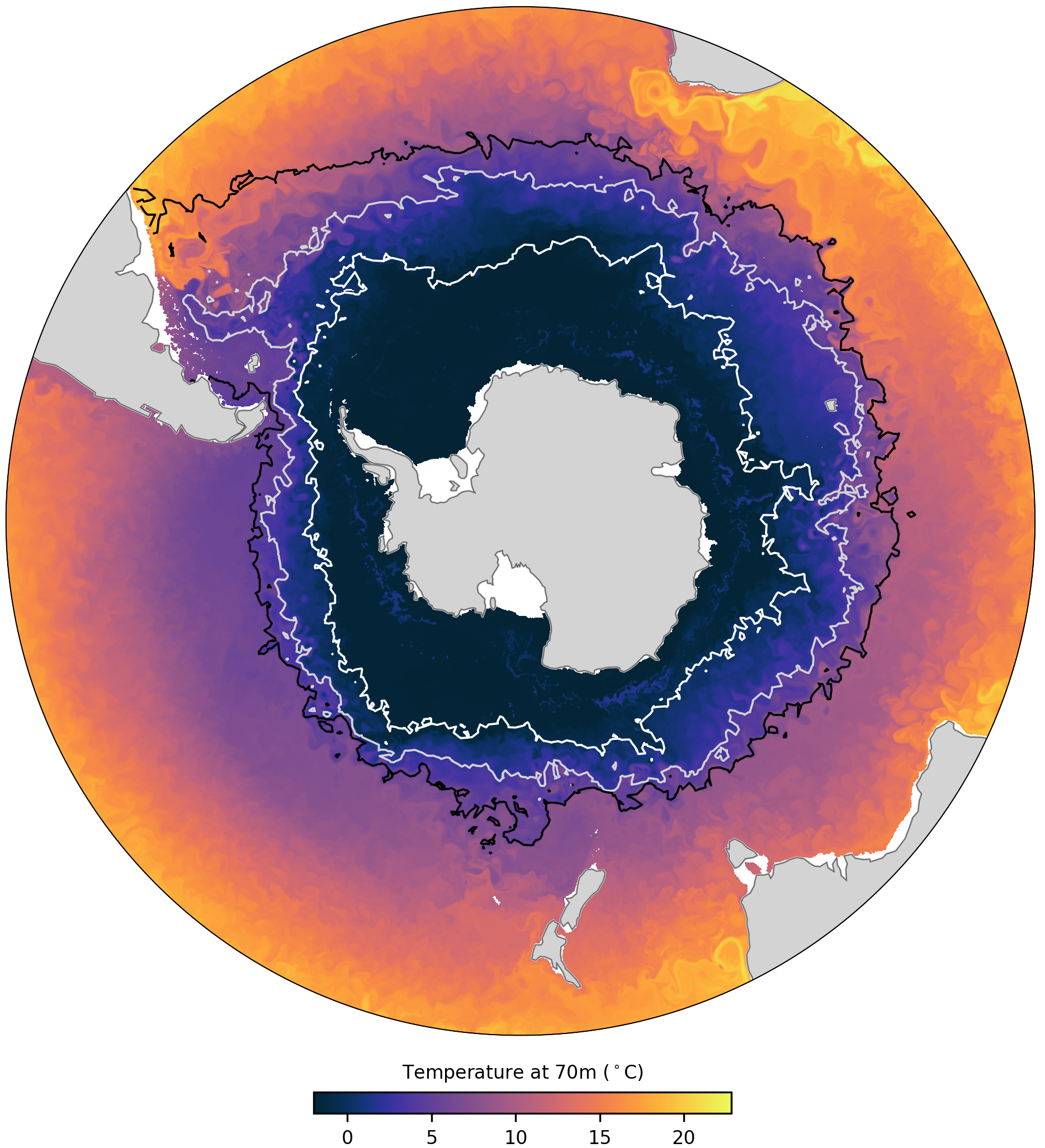}
\caption{This figure is similar to Figure \ref{fig:fronts_fig1}, except we are showing the full conservative temperature field rather than only on a segment of the polar domain. Three concentric circumpolar fronts are identified as material contours that move with the flow while minimizing deformation and mixing over a 540-day time window (Methods). Each front is plotted at the center of the time window (day 270).
The fronts are computed using simulated particle trajectories at fixed depth (70m) calculated from a 0.1$^\circ$ global eddy-resolving ocean model. Background color shows a snapshot (on day 270) of daily averaged conservative temperature at 70m. The identified fronts partition the Southern Ocean into sub-domains of different water mass properties, often aligning with steep gradients in the conservative temperature field.}
\label{fig:supp_fronts_temp}
\end{figure}

\begin{figure}[H]
\centering \includegraphics[width=0.8\linewidth]{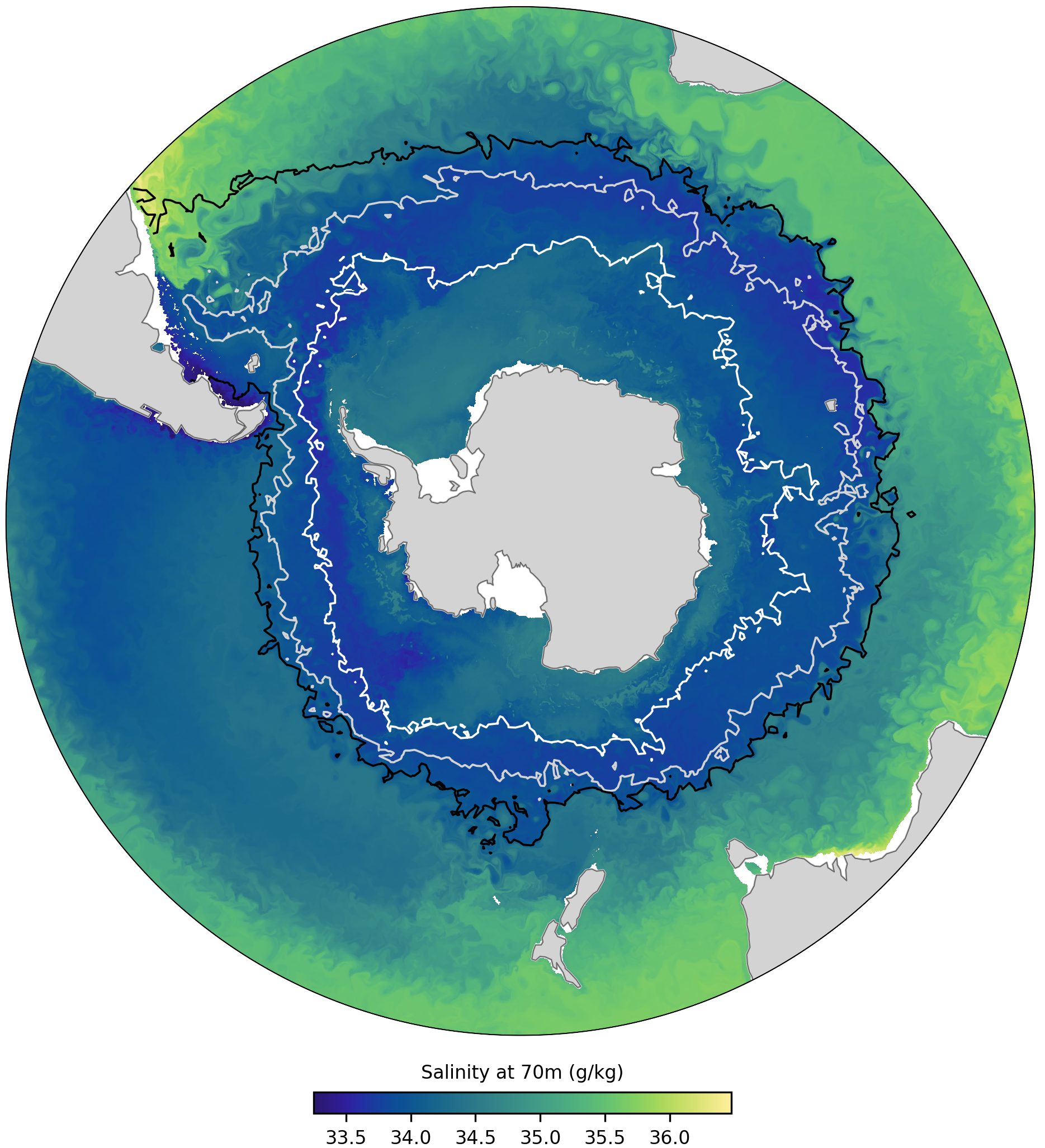}
\caption{This figure is similar to Figure \ref{fig:fronts_fig1}, except we are showing the full practical salinity field rather than only on a segment of the polar domain. Three concentric circumpolar fronts are identified as material contours that move with the flow while minimizing deformation and mixing over a 540-day time window (Methods). Each front is plotted at the center of the time window (day 270).
The fronts are computed using simulated particle trajectories at fixed depth (70m) calculated from a 0.1$^\circ$ global eddy-resolving ocean model. Background color shows a snapshot (on day 270) of daily averaged practical salinity at 70m. The identified fronts partition the Southern Ocean into sub-domains of different water mass properties, often aligning with steep gradients in the practical salinity field.}
\label{fig:supp_fronts_salt}
\end{figure}

\begin{figure}[H]
\centering \includegraphics[width=0.8\linewidth]{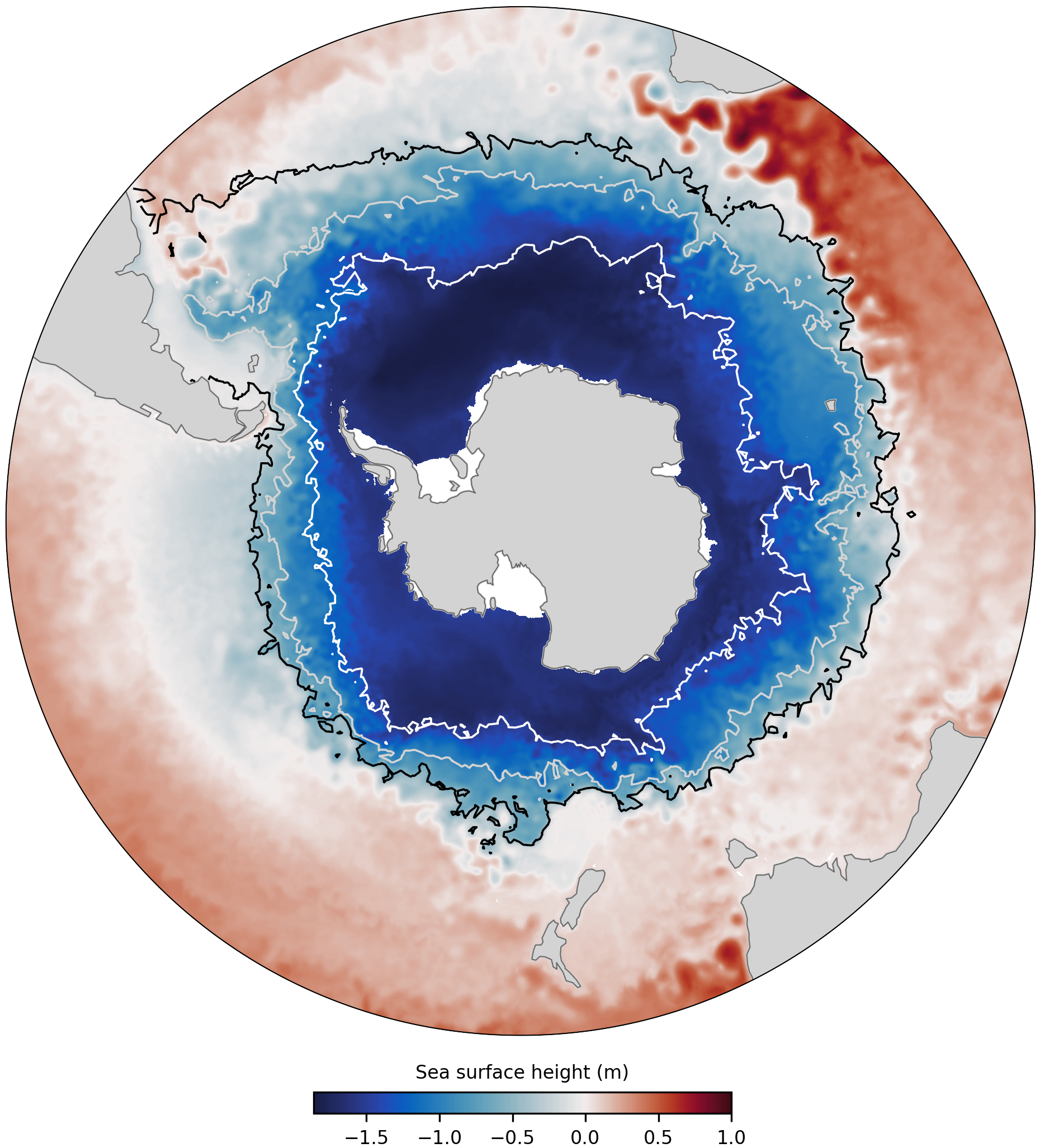}
\caption{This figure is similar to Figure \ref{fig:fronts_fig1}, except we are showing the full effective sea surface height field rather than only on a segment of the polar domain. Three concentric circumpolar fronts are identified as material contours that move with the flow while minimizing deformation and mixing over a 540-day time window (Methods). Each front is plotted at the center of the time window (day 270).
The fronts are computed using simulated particle trajectories at fixed depth (70m) calculated from a 0.1$^\circ$ global eddy-resolving ocean model. Background color shows a snapshot (on day 270) of daily averaged effective sea surface height at 70m. The identified fronts partition the Southern Ocean into sub-domains of different water mass properties, often aligning with steep gradients in the effective sea surface height field.}
\label{fig:supp_fronts_ssh}
\end{figure}

\begin{figure}[H]
\centering
\includegraphics[width=\linewidth]{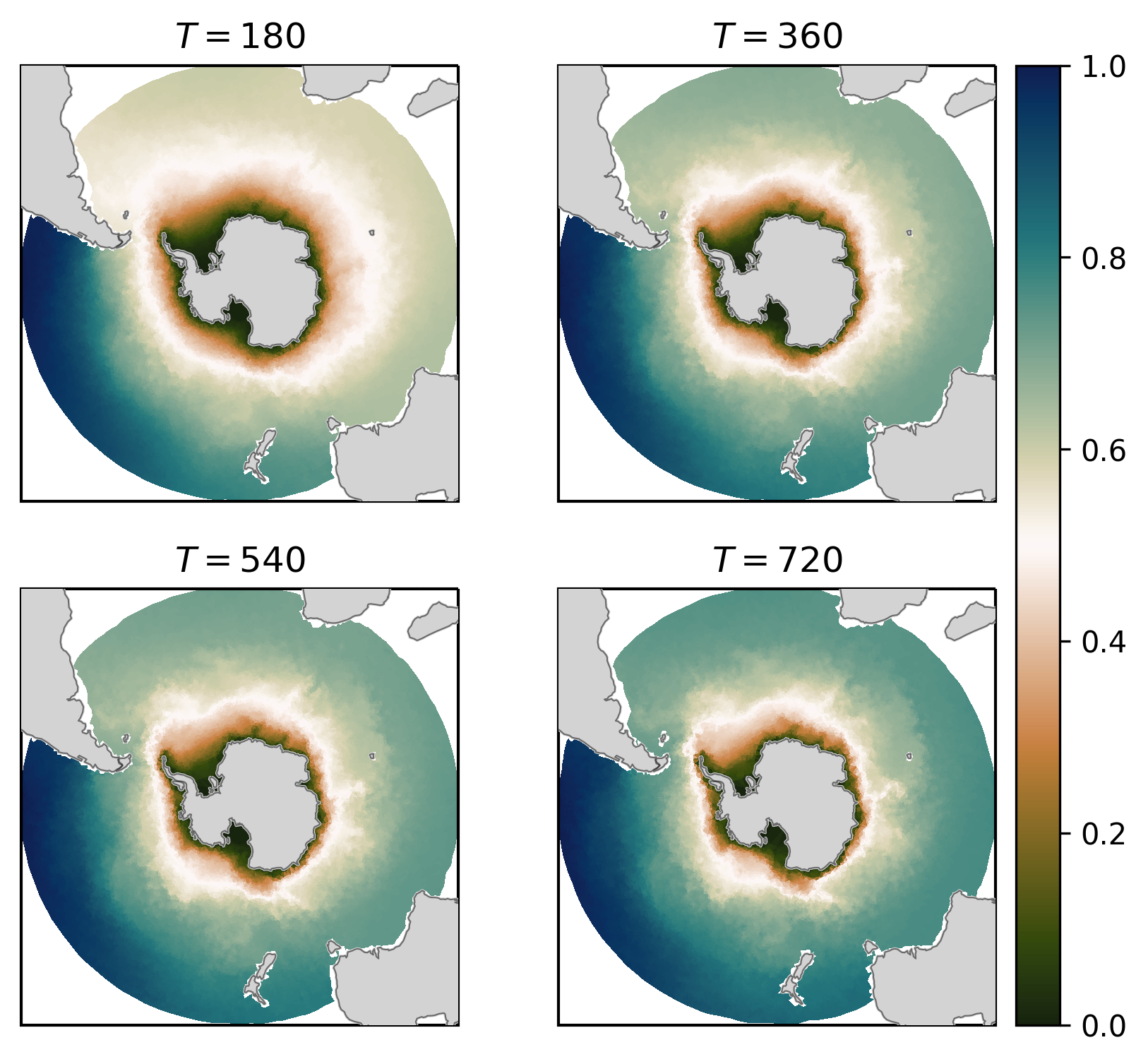}
\caption{The dominant eigenfunction of $\Delta^{D}_{W(365,T)}$ for window lengths $T=180$ (top left), $T=360$ (top right), $T=540$ (bottom left), and $T=720$ (bottom right). There is little discernible difference between the structure of the eigenfunctions for lengths $T=540$ and $T=720$. }
\label{fig:telescoping_eigenfunctions}
\end{figure}

\begin{figure}[H]
\centering \includegraphics[width=\linewidth]{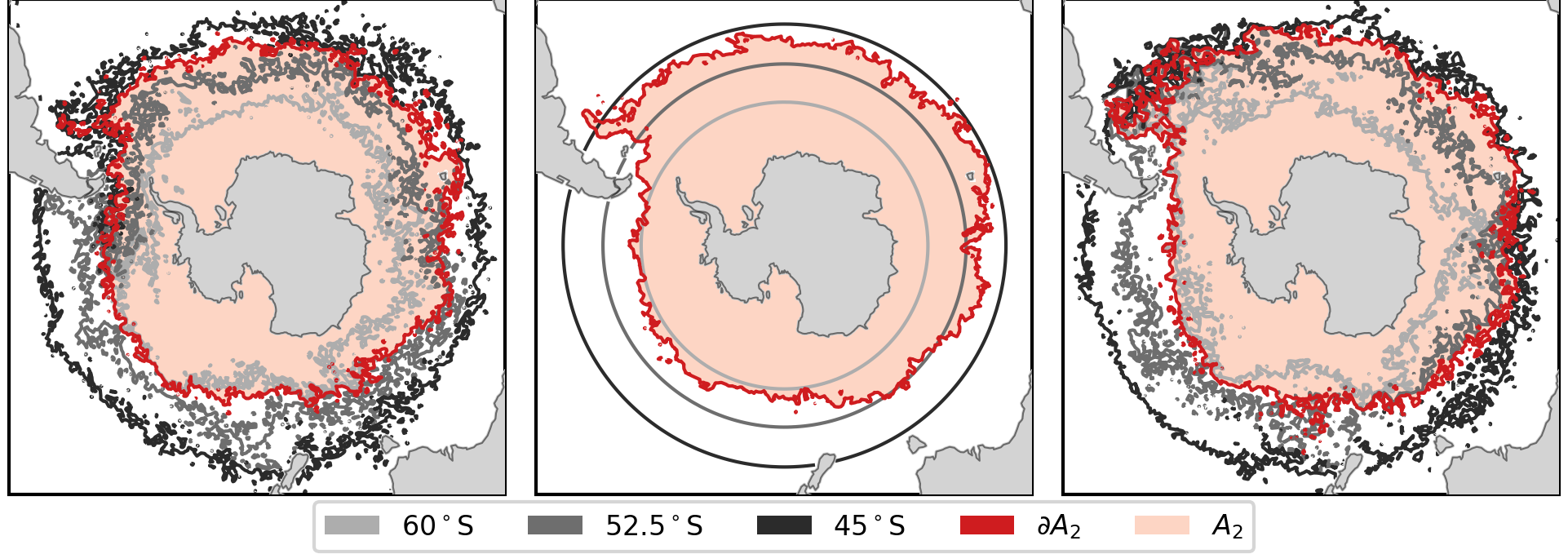}
\caption{\textbf{The resistance of fronts to distortion and filamentation.} The middle front $\partial A_{2}$ (red curve) identified in Figure \ref{fig:fronts_fig1}, and the finite-time coherent set $A_{2}$ (red shaded region) are plotted in the center panel at the center of their window $W(270,540)$ (observation day 270), along with their 270-day backward evolution (left panel) and 270-day forward evolution (right). Plotted in shades of gray are three latitude rings defined in the center panel on observation day 270, along with their 270-day backward evolution (left), and 270-day forward evolution (right). The red front experiences significantly less distortion and filamentation compared to the three gray rings of latitude. As small-scale diffusive mixing between two regions can only occur along their interface, the identified front acts a barrier to this transport as its evolved length remains relatively unchanged.} \label{fig:fronts_ftcs_diagram}
\end{figure}

\paragraph{Southern Ocean fronts in regional domains:}
In the main text, we defined circumpolar Southern Ocean fronts as circumpolar material curves that are minimally stretched and distorted by the flow, and identified the fronts with boundaries of finite-time coherent sets. We chose the domain $\mathcal{M}$ as the region of ocean south of $30^\circ$S, where we applied Dirichlet boundary conditions on the Antarctic coastline, and Neumann boundary conditions on the remainder of the domain boundary. In this section we compare these circumpolar Southern Ocean fronts to Southern Ocean fronts identified on regional domains.

We now denote by $\mathcal{M}$ a region of the Southern Ocean at 70m depth (with land removed). The boundary $\partial \mathcal{M}$ may be  open ocean and/or coastlines. With Neumann boundary conditions imposed on $\partial \mathcal{M}$, we seek a smooth curve $\Gamma$ that disconnects $\mathcal{M}$ in such a way that the average evolved length of $\Gamma$ is minimal relative to the volume of the two disconnected pieces over a prescribed window $W(t_c,T)$ (see \cite{Froyland2018}). Here, $\Gamma$ describes a transport barrier in the domain $\mathcal{M}$, which we define as a \emph{regional Southern Ocean front} in the domain $\mathcal{M}$. The regional front $\Gamma$ can be approximated by a level set of the leading non-trivial eigenfunction of a dynamic Laplace operator \cite{Froyland2015a}, defined in Eq. (\ref{eq:dynamic_laplace}), with Neumann boundary conditions
\begin{align*}\label{eq:mixed_eigenproblem}
\Delta^D_{W(0,T)} v &= \lambda v \quad \text{ on int}(\mathcal{M}),\\
\int_{-T/2}^{T/2} \frac{\partial \Phi^{t}_{*} v}{\partial n^{t}}\circ \Phi^{t}\text{ d}t &= 0 \quad \text{ on } \partial \mathcal{M}.
\end{align*}

To compare the circumpolar Southern Ocean fronts to regional fronts, we selected seven regions of interest;
$R_1 = [30^\circ\text{W}, 30^\circ{E}]   \times[65^\circ\text{S}, 38^\circ\text{S}]$,
$R_2 = [30^\circ\text{E}, 100^\circ{E}]  \times[65^\circ\text{S}, 35^\circ\text{S}]$,
$R_3 = [100^\circ\text{E}, 145^\circ{E}] \times[65^\circ\text{S}, 40^\circ\text{S}]$,
$R_4 = [145^\circ\text{E}, 170^\circ{W}] \times[65^\circ\text{S}, 42^\circ\text{S}]$,
$R_5 = [170^\circ\text{W}, 105^\circ{E}]\times[70^\circ\text{S}, 45^\circ\text{S}]$,
$R_6 = [105^\circ\text{W}, 68^\circ{W}]\times[67^\circ\text{S}, 50^\circ\text{S}]$,
$R_7 = [68^\circ\text{W}, 30^\circ{W}]\times[65^\circ\text{S}, 30^\circ\text{S}]$.
These regions were selected due to the unique bathymetry and ACC flow characteristics within each region. We focus on the window $W(270,540)$. Given the long window length, a large proportion of particles that start in a given region will exit that region relatively quickly. To avoid tracking many exiting particles we simply fix a given region $R$ (one of $R_1,\ldots,R_7$) and set $\mathcal{M}_{t}=R$ at time $t$;  thus we allow particles to enter and/or exit $\mathcal{M}_t$ during the window $W(270,540)$. For times where a particle is outside of the region, we treat that particle as missing. In these experiments we used a higher density particle set ($400,000$ particles initialized on a Fibonacci lattice rather than $60,000$ as was used in the main text) advected using the same model described in the main text. We used the same numerical implementation described in the main text, ignoring Dirichlet boundary conditions.

\begin{figure}[H]
\centering \includegraphics[width=\linewidth]{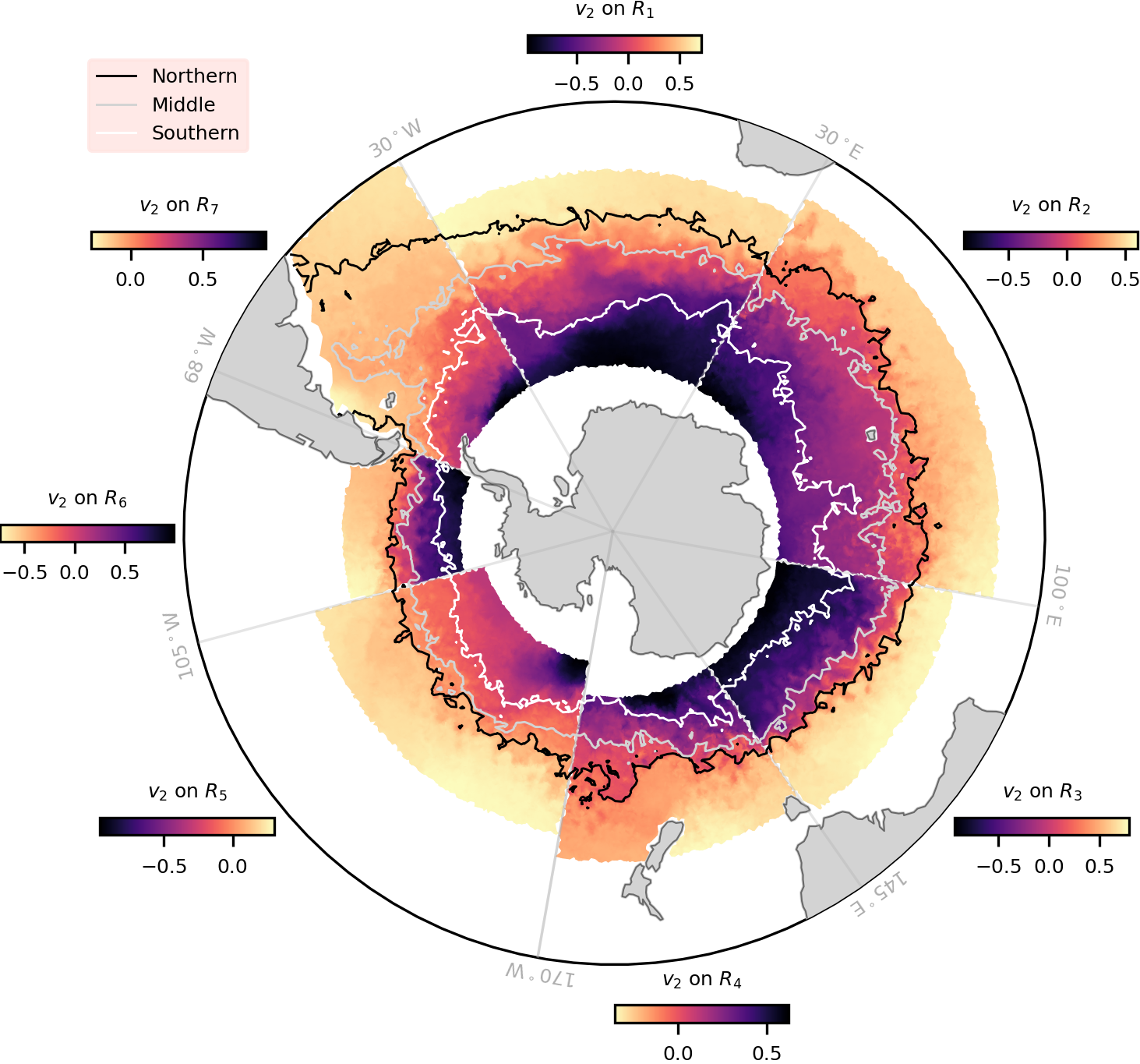}
\caption{\textbf{Circumpolar Southern Ocean fronts overlaid on functions that describe regional Southern Ocean fronts.} Three concentric circumpolar fronts are identified as material contours that move with the flow while minimizing deformation and mixing over a 540-day time window (Methods). Each front is plotted at the center of the time window (day 270).
The fronts are computed using simulated particle trajectories at fixed depth (70m) calculated from a 0.1$^\circ$ global eddy-resolving ocean model. Background colors show the dominant non-trivial eigenfunctions of dynamic Laplace operators ($v_2$) defined on each regional domain using Neumann boundary conditions. The circumpolar fronts generally align well with contours of the dominant non-trivial eigenfunctions, particularly in locations away from the boundaries of the domains.} \label{fig:supp_local_fronts}
\end{figure}

In Figure \ref{fig:supp_local_fronts} we plot the three circumpolar fronts identified in Figure \ref{fig:fronts_fig1}, overlaid on the leading non-trivial eigenfunctions of dynamic Laplace operators constructed for the seven regions of interest.
The circumpolar fronts generally align well with contours of the dominant non-trivial eigenfunctions (which define regional fronts), particularly in locations away from the boundaries of the domains.
This demonstrates that our front extraction approach is robust to changes in implementation, including using smaller domains, using a fixed domain rather than a mobile domain, and differing boundary conditions.

\paragraph{Animation of the time series of fronts:}

An animation of the time series of fronts identified from a sequence of sliding windows can be found at \href{https://youtu.be/e0GW1gd1DgY}{https://youtu.be/e0GW1gd1DgY}.

\begin{figure}[H]
\centering \includegraphics[width=\linewidth]{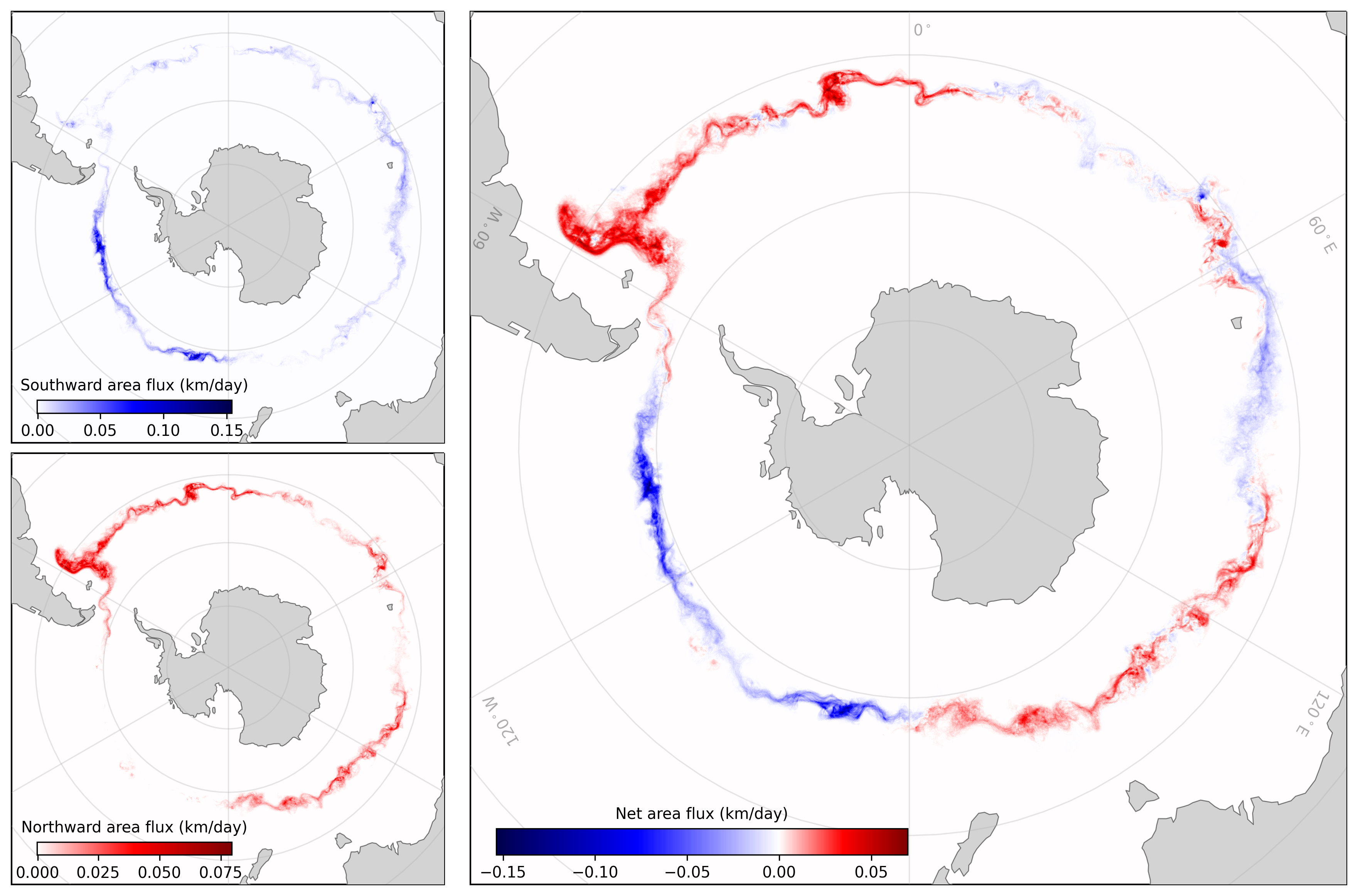}
\caption{This figure is similar to Figure \ref{fig:fronts_fig3}, except the temporal average is taken over the first year of observations only. A 1-year sequence of daily front locations is constructed by tracking the middle front identified from Figure \ref{fig:fronts_fig1}. For each center time, the northward (red) and southward (blue) cross-front transport regions are identified. A temporal average of these regions normalized by the contour length is computed on a $0.1^\circ \times 0.1^\circ$ uniform grid in order to compute an area flux. The average southward cross-front area flux is plotted in the top left panel, and the average northward cross-front area flux in the bottom left panel. The average net cross-front transport is computed as the difference of these two fields, show in the right panel. A distinctive global pattern of net cross-front transport emerges, where the net cross-front transport transitions from northward to southward, or southward to northward, downstream of prominent bathymetric obstacles. After each transition we see particularly large amounts of total cross-front transport.}
\label{fig:supp_fronts_fig3_1yr}
\end{figure}

\begin{figure}[H]
\centering \includegraphics[width=\linewidth]{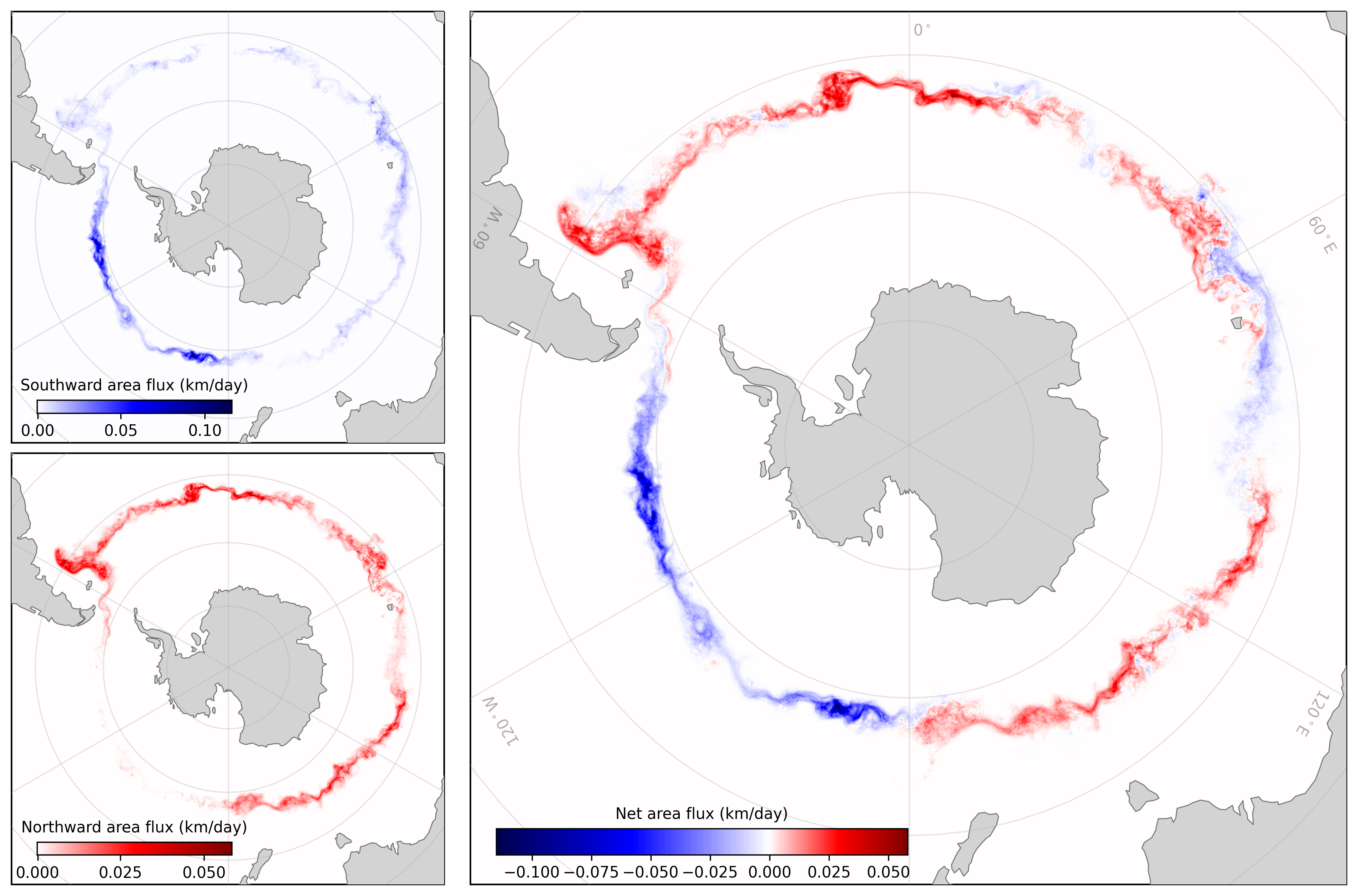}
\caption{This figure is similar to Figure \ref{fig:fronts_fig3}, except the temporal average is taken over the first two years of observations only. A 2-year sequence of daily front locations is constructed by tracking the middle front identified from Figure \ref{fig:fronts_fig1}. For each center time, the northward (red) and southward (blue) cross-front transport regions are identified. A temporal average of these regions normalized by the contour length is computed on a $0.1^\circ \times 0.1^\circ$ uniform grid in order to compute an area flux. The average southward cross-front area flux is plotted in the top left panel, and the average northward cross-front area flux in the bottom left panel. The average net cross-front transport is computed as the difference of these two fields, show in the right panel. A distinctive global pattern of net cross-front transport emerges, where the net cross-front transport transitions from northward to southward, or southward to northward, downstream of prominent bathymetric obstacles. After each transition we see particularly large amounts of total cross-front transport.}
\label{fig:supp_fronts_fig3_2yr}
\end{figure}

\begin{figure}[H]
\centering \includegraphics[width=\linewidth]{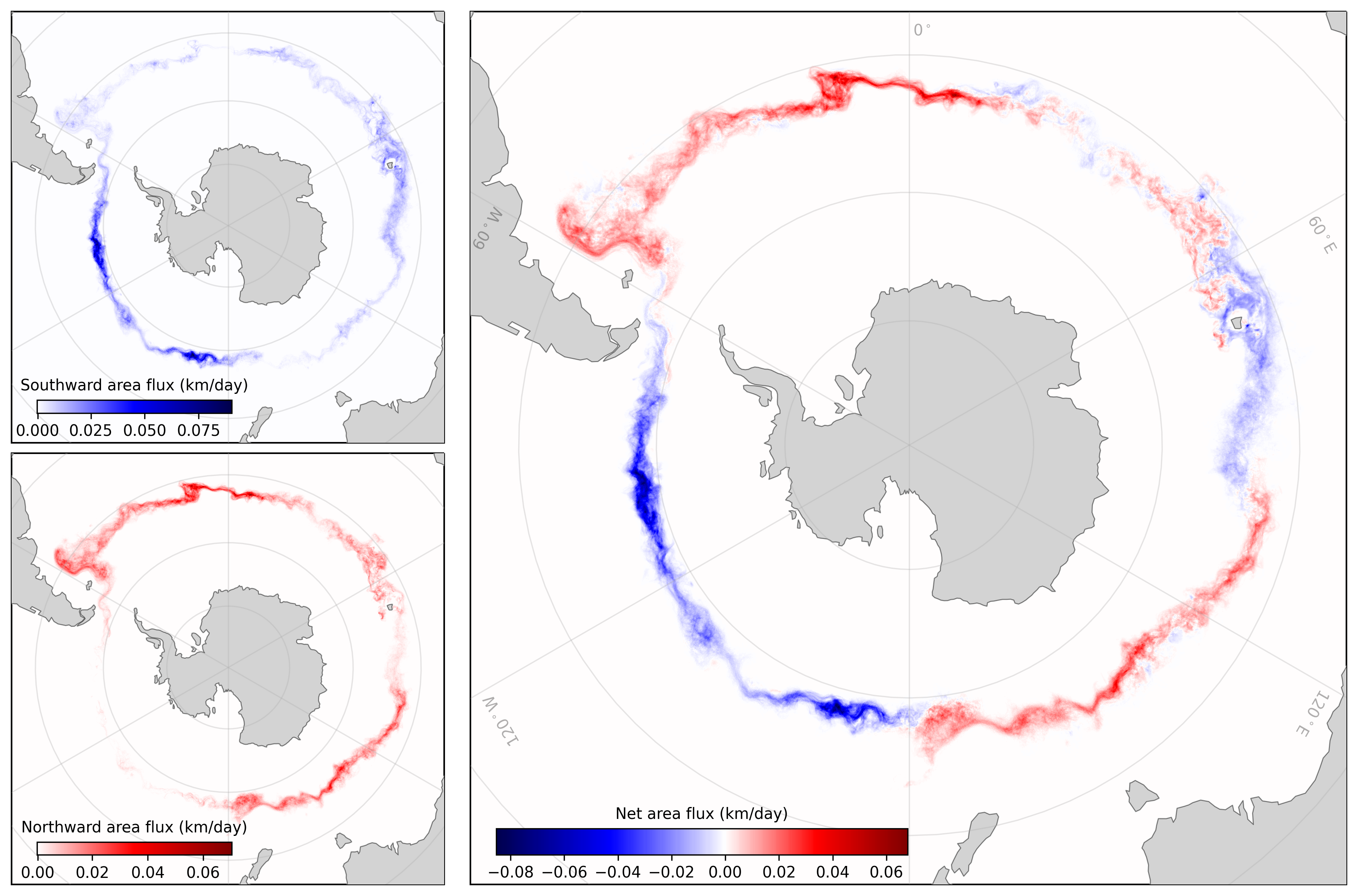}
\caption{This figure is similar to Figure \ref{fig:fronts_fig3}, except the temporal average is taken over the first three years of observations only. A 3-year sequence of daily front locations is constructed by tracking the middle front identified from Figure \ref{fig:fronts_fig1}. For each center time, the northward (red) and southward (blue) cross-front transport regions are identified. A temporal average of these regions normalized by the contour length is computed on a $0.1^\circ \times 0.1^\circ$ uniform grid in order to compute an area flux. The average southward cross-front area flux is plotted in the top left panel, and the average northward cross-front area flux in the bottom left panel. The average net cross-front transport is computed as the difference of these two fields, show in the right panel. A distinctive global pattern of net cross-front transport emerges, where the net cross-front transport transitions from northward to southward, or southward to northward, downstream of prominent bathymetric obstacles. After each transition we see particularly large amounts of total cross-front transport.}
\label{fig:supp_fronts_fig3_3yr}
\end{figure}

\begin{figure}[H]
\centering
\includegraphics[width=\linewidth]{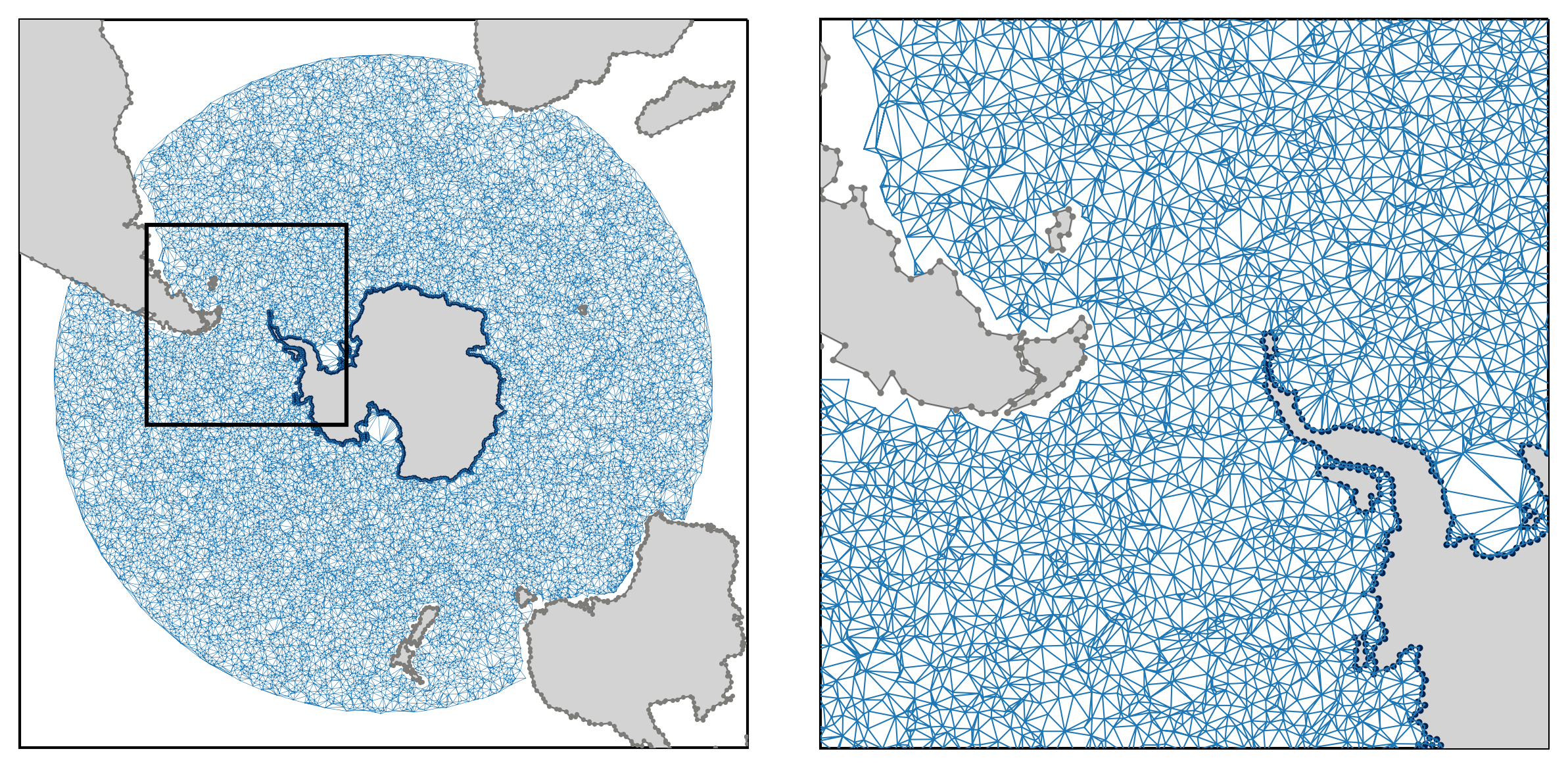}
\caption{The mesh of $\mathcal{M}_{270}$ represented in a South Polar Stereographic projection, with Antarctic coastline points plotted in dark blue and non-Antarctic coastline points plotted in gray.}
\label{fig:mesh}
\end{figure}

\end{document}